\documentclass[aps, twocolumn, amsmath, amssymb, prb, 10pt, floatfix]{revtex4-1}

\usepackage[per-mode=symbol,separate-uncertainty]{siunitx}
\DeclareSIUnit \belm {Bm}

\usepackage[]{graphics} 
\usepackage{amsmath,color,colordvi}  
\usepackage{mathtools}
\usepackage[pdftex]{graphicx} 
\usepackage[english]{babel}   
\usepackage{braket}
 
\definecolor{navyblue}{rgb}{0.0, 0.0, 0.5}
\newcommand{\Dn}[0]{{\rm d}}
\newcommand{\ii}[0]{{\rm i}}
\newcommand{\ee}[0]{{\rm e}}

\newcommand{\normrateright}[0]{\overrightarrow{F}}

\newcommand{\nmd}[0]{\rm N}

\usepackage[protrusion=true, expansion=true]{microtype} 
\usepackage[colorlinks=true, breaklinks=false, linkcolor=navyblue, urlcolor=navyblue, citecolor=navyblue]{hyperref}
 
\begin{document}

\title{Observation of a broadband Lamb shift in an engineered quantum system}

\author{Matti Silveri$^{1,2}$}
\email{matti.silveri@oulu.fi}
\author{Shumpei Masuda$^{1,3}$}
\author{Vasilii Sevriuk$^1$}
\author{Kuan Y. Tan$^1$}
\author{M\'at\'e Jenei$^1$}
\author{Eric Hyypp\"a$^1$}
\author{Fabian~Hassler$^4$}
\author{Matti~Partanen$^1$}
\author{Jan Goetz$^1$}
\author{Russell E. Lake$^{1,5}$}
\author{Leif Gr\"onberg$^6$}
\author{Mikko M\"{o}tt\"{o}nen$^1$}
\email{mikko.mottonen@aalto.fi}

\affiliation{$^1$QCD Labs, QTF Center of Excellence, Department of Applied Physics, Aalto University, P.O. Box 13500, FI-00076 Aalto, Finland\\
  $^2$Research Unit of Nano and Molecular Systems, University of Oulu, P.O. Box 3000, FI-90014 Oulu, Finland\\
  $^3$Collage of Liberal Arts and Sciences, Tokyo Medical and Dental University, Ichikawa, 272-0827, Japan\\
  $^4$JARA Institute for Quantum Information, RWTH Aachen University, 52056 Aachen, Germany\\
  $^5$National Institute of Standards and Technology, Boulder, Colorado 80305, USA\\$^6$VTT Technical Research Centre of Finland, QTF Center of Excellence, P.O. Box 1000, FI-02044 VTT, Finland}
 
\date{\today}

\maketitle    
\textbf{The shift of energy levels owing to broadband electromagnetic vacuum fluctuations\----the Lamb shift\----has been pivotal in the development of quantum electrodynamics and in understanding atomic spectra~\cite{Lamb47, Bethe47, Heinzen87, Brune94, Marrocco98,  Carmichael}. Currently, small energy shifts in engineered quantum systems are of paramount importance owing to the extreme precision requirements in applications such as quantum computing~\cite{Gisin07, Ladd10}. However, without a tunable environment it is  challenging to resolve the Lamb shift in its original broadband case. Consequently, the observations in other than atomic systems~\cite{Lamb47, Bethe47, Heinzen87, Brune94, Marrocco98, Rentrop16} are limited to environments comprised of narrow-band modes~\cite{Fragner08, Yoshihara18, Mirhosseini18}. Here, we observe a broadband Lamb shift in high-quality superconducting resonators, a scenario also accessing  any static shift inaccessible in Lamb's experiment~\cite{Lamb47, Bethe47}. We measure a continuous change of several megahertz in the fundamental resonator frequency by externally tuning the coupling strength of the engineered broadband environment which is based on hybrid normal-metal--superconductor tunnel junctions~\cite{Partanen16, Tan16, Masuda16}. Our results may lead to improved  control of dissipation in high-quality engineered quantum systems and open new possibilities for studying synthetic open quantum matter~\cite{Houck12, Fitzpatrick17, Ma18} using this hybrid experimental platform.}

\begin{figure*}
  \begin{center}
    \includegraphics[width=1\linewidth]{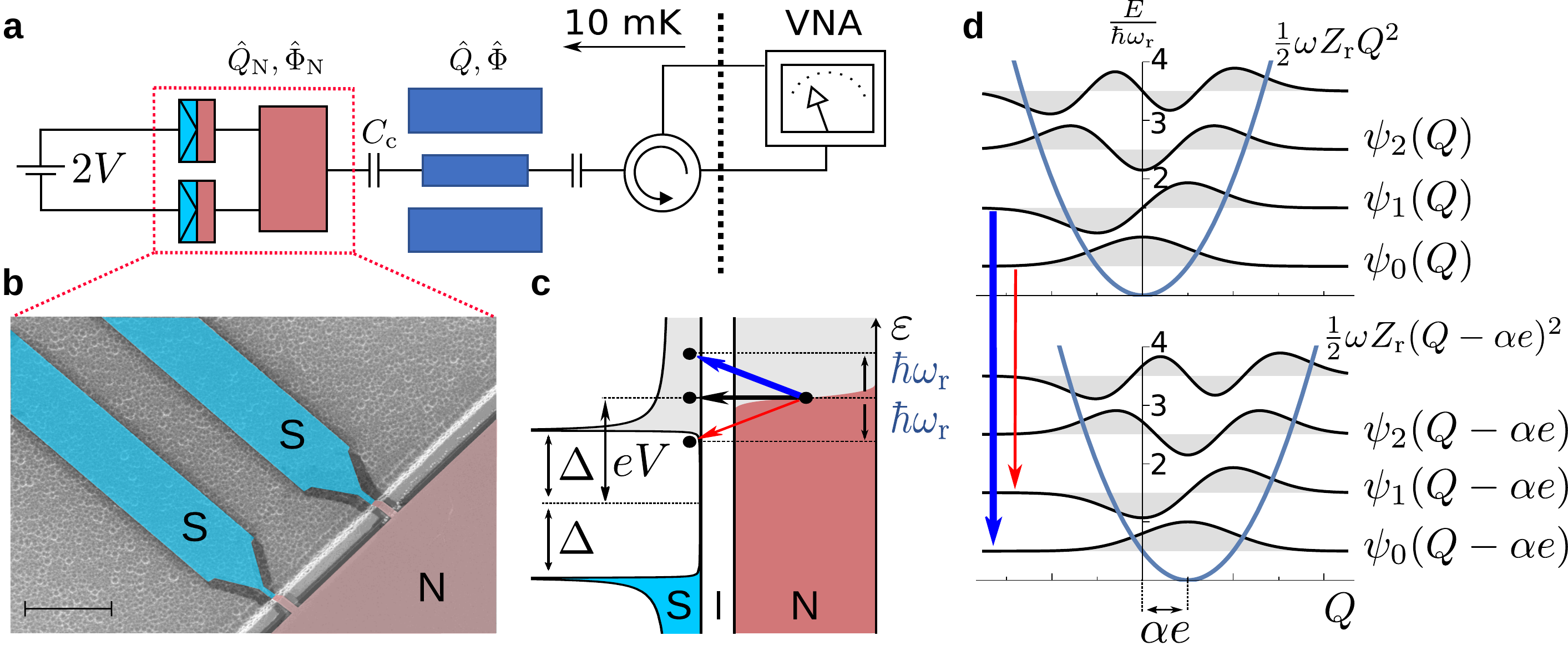}     
  \end{center}
  \caption{\label{fig:teaser} \textbf{Sample and measurement setup.}  \textbf{a},~Schematic illustration of the coplanar waveguide resonator (dark blue) capacitively coupled to a normal-metal island (red) and a transmission line together with a simplified measurement setup. \textbf{b},~False-colour scanning electron microscope image of the two superconductor--insulator--normal-metal~(SIN) tunnel junctions used as an engineered environment for the resonator modes. The scale bar denotes \SI{5}{\micro\meter}. See Supplementary Figs.~S1--S2 for details of the sample and the measurement setup. \textbf{c},~Energy diagram of photon-assisted tunneling at a superconductor--insulator--normal-metal junction. In the normal-metal, the electron occupation~(red shading) follows the Fermi distribution. The superconductor density of states exhibits the characteristic Bardeen--Cooper--Schrieffer energy gap of magnitude~$2\Delta$. The states below the gap are filled~(blue shading). The gray shading denotes empty states. The blue arrow depicts a tunneling process that absorbs a photon with energy~$\hbar \omega_{\rm r}$ from the resonator mode at the angular frequency~$\omega_{\rm r}$. The red arrow corresponds to photon emission. Elastic processes (black arrow) do not affect the resonator state but contribute to the Lamb shift and to the thermalization of the normal-metal island~\cite{Silveri17prb}. The bias voltage $V$ shifts the electrochemical potentials of the normal metal and of the superconductor relative to each other by $eV$. For voltage biases $|eV|<\Delta+\hbar\omega_{\rm r}$, emission processes are suppressed by the vanishing density of states in the superconductor gap. \textbf{d},~A~tunneling event on the normal-metal island shifts the charge of the resonator by $\Delta Q=\alpha e$. The capacitance fraction $\alpha = C_{\rm c}/(C_{\rm c}+C_{\Sigma m})\approx 1 $ is given by the coupling capacitance $C_{\rm c}$ between the resonator and the normal-metal island and the capacitance of the normal-metal island to ground $C_{\rm \Sigma m}$~(Table~\ref{tab:parameters}). The charge shift induces transitions between the resonator energy eigenstates $\psi_i(Q)$ and $\psi_f(Q)$ via the matrix element $|M_{if}|^2=|\int\psi_f^\ast (Q-\alpha e) \psi_i(Q)  \Dn{Q}|^2 \propto \rho^{|i-f|}$, where $\rho=\pi \alpha^2 Z_{\rm r}/R_{\rm K}$ is an interaction parameter expressed in terms of the characteristic impedance $Z_{\rm r}$ of the resonator and the von Klitzing constant $R_{\rm K}=h/e^2$ containing the Planck constant~$h$~(Methods). The blue and red arrows correspond to those in~\textbf{c}. }
\end{figure*}

Physical quantum systems are always open. Thus, exchange of energy and information with an environment eventually leads to relaxation and degradation of quantum coherence. Interestingly, the environment can be in a vacuum state and yet cause significant perturbation to the original quantum system. The quantum vacuum can be modelled as broadband fluctuations which may absorb energy from the coupled quantum systems. These fluctuations also lead to an energy level renormalization\----the Lamb shift\----of the system, such as that observed in atomic systems~\cite{Lamb47, Bethe47, Heinzen87, Brune94, Marrocco98, Rentrop16}. Despite of its fundamental nature, the  Lamb shift arising from broadband fluctuations is often overlooked outside the field of atomic physics as a small constant shift that is challenging to distinguish~\cite{Gramich2011}. Due to the emergence of modern engineered quantum systems, in which the desired precision of the energy levels is comparable to the Lamb shift, it has, however, become important to predict accurately the perturbation as a function of external control parameters. Neglecting energy shifts can potentially take the engineered quantum systems outside the region of efficient operation~\cite{Paraoanu06, Rigetti10} and may even lead to undesired level crossings between subsystems. These issues are pronounced in applications requiring strong dissipation. Examples include reservoir engineering for autonomous quantum error correction~\cite{Kerckhoff10, Kapit15}, or rapid on-demand entropy and heat evacuation~\cite{Geerlings13, Tan16, Masuda16, Partanen18}.  Furthermore, the role of dissipation in phase transitions of open many-body quantum systems has attracted great interest through the recent progress in studying synthetic quantum matter~\cite{Houck12, Fitzpatrick17}. 

In our experimental setup, the system exhibiting the Lamb shift is a superconducting coplanar waveguide resonator with the resonance frequency $\omega_{\rm r}/2\pi=$~\SI{4.7}{\giga\hertz} and \SI{8.5}{\giga\hertz} for Sample A and B, respectively, with loaded quality factors in the range of \numrange{e2}{e3}. The total Lamb shift includes two parts: the dynamic part~\cite{Weiss, Bethe47, Frisk14} arising from the fluctuations of the broadband electromagnetic environment formed by electron tunneling across normal-metal--insulator--superconductor junctions~\cite{IngoldNazarov05, Tan16, Silveri17prb, Masuda16}~(Fig.~\ref{fig:teaser}) and the static shift originating here from the environment-induced change of the resonator mode. Our system differs in three key ways from the Lamb shift typically observed in atoms coupled to electromagnetic radiation~\cite{Lamb47, Bethe47, Heinzen87, Brune94, Marrocco98, Rentrop16}. First, in our case an electron system induces a frequency shift to the electromagnetic system and not vice versa as for atoms. Second, we can access the system also when it is essentially decoupled from the environment in contrast to the typical case of an atom where the electrons are always coupled to the electromagnetic environment. Third, our system is sensitive to both the static and the dynamic part of the Lamb shift. This is a striking difference to atomic systems, where the static part is typically inaccessible since it corresponds to the additional electromagnetic mass already included in the measured masses of the particles. 

\begin{figure*}
  \centering 
  \includegraphics[width=1\linewidth]{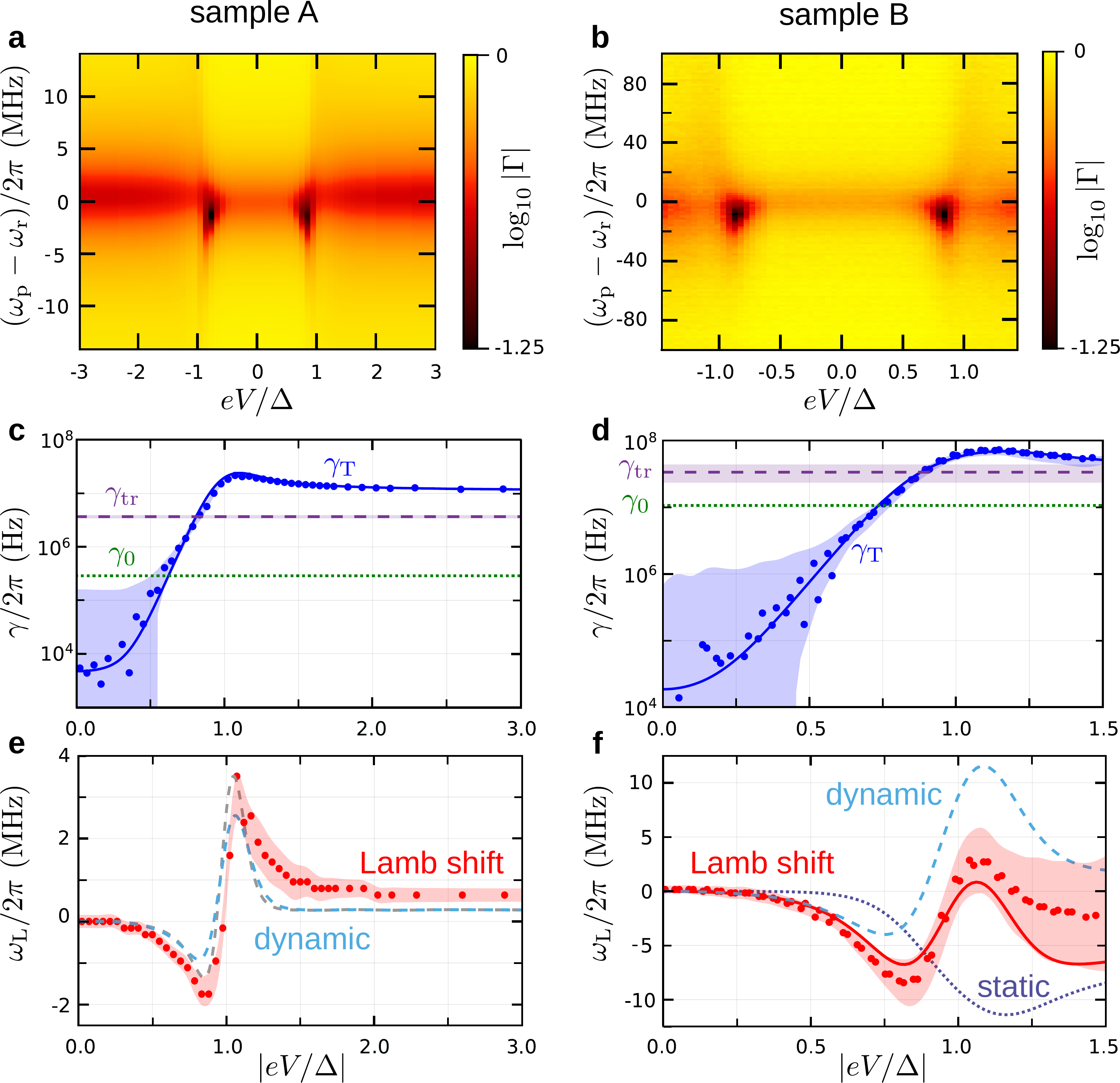}
  \caption{\label{fig:GammaLamb} \textbf{Observation of the Lamb shift.} ~\textbf{a, b}, Magnitude of the voltage reflection coefficient $|\Gamma|$ as a function of the probe frequency $\omega_{\rm p}$ and of the single-junction bias voltage $V$. \textbf{c, d},~Coupling strength $\gamma_{\rm T}$ to the electromagnetic environment formed by the photon-assisted tunneling at the superconductor--insulator--normal-metal junctions as a function of the the single-junction bias voltage $V$. For the calculated coupling strengths~(solid lines) we use the experimentally realized parameter values, see Table~\ref{tab:parameters}. The horizontal dashed lines denote the coupling strength to the transmission line $\gamma_{\rm tr}$ and the horizontal dotted lines indicate the coupling strength to excess sources $\gamma_0$. \textbf{e, f},~The~Lamb~shift as a function of the single-junction bias voltage $V$~(solid circles). The solid line of the panel \textbf{f} denotes the total Lamb shift including both the static~(dotted line) and the dynamic~(dashed line) parts. The gray dashed line in panel~\textbf{e} shows the dynamic Lamb shift corresponding the electron temperature $T_{\rm N}=$~\SI{130}{\milli\kelvin}. Panels \textbf{a}, \textbf{c}, and \textbf{e} are for Sample A and \textbf{b}, \textbf{d}, and \textbf{f} correspond to Sample B. The shaded regions denote the $1\sigma$ confidence intervals of the extracted parameters (see Methods). The excess coupling strenght $\gamma_0$ has a similar confidence interval (not shown) than the the coupling strength to the transmission line $\gamma_{\rm tr}$.}
\end{figure*}

\begin{figure}
  \centering  
   \includegraphics[width=1.0\linewidth 
   ]{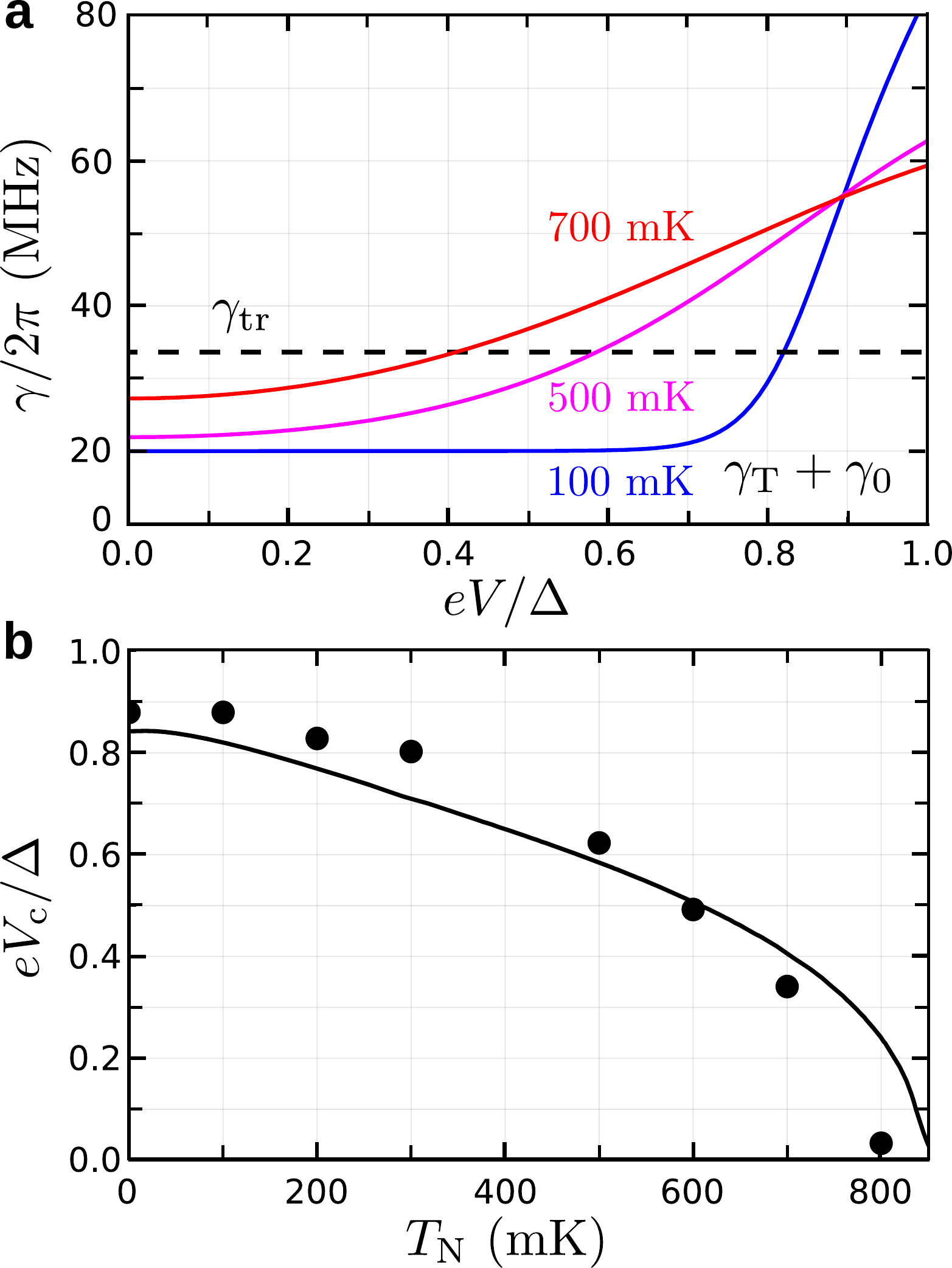} 
  \caption{\label{fig:T} \textbf{Temperature dependence.} \textbf{a},~The calculated total coupling strength $\gamma_{\rm T}+\gamma_0$ as a function of the single-junction bias voltage at the normal-metal electron temperature $T_{\rm N}=$\SI{100}{\milli\kelvin}~(blue), \SI{500}{\milli\kelvin}~(magenta), and \SI{700}{\milli\kelvin}~(red) with parameters of Sample B~(Table~\ref{tab:parameters}). The horizontal dashed line indicates the coupling strength of the transmission line $\gamma_{\rm tr}$. The coincidence point $\gamma_{\rm T}+\gamma_0=\gamma_{\rm tr}$ defines the critical bias value $V_{\rm c}$, where the reflection coefficient ideally vanishes.  The single-junction bias voltage is measured in the units of the zero-temperature superconductor gap $\Delta/e$ and the theoretical calculation takes into account the temperature dependence of the gap. \textbf{b}, The critical voltage $V_{\rm c}$ as a function of the normal-metal electron temperature $T_{\rm N}$ for Sample B. The data points~(filled circles) correspond to the bias voltage of the minima of the measured voltage reflection coefficients (Supplementary Fig.~S5). For the calculated critical voltage~(solid line) we use experimentally realized parameters~(Table~\ref{tab:parameters}) except that the value of the excess coupling strength is $\gamma_0/2\pi=$\SI{20.0}{\mega\hertz} capturing the enhanced losses by excess quasiparticles in the superconducting coplanar waveguide resonator at high temperatures.
  }
\end{figure}

We observe that the coupling strength between the environment and the resonator $\gamma_{\rm T}/2\pi $ can be tuned from \SI{10}{\kilo\hertz}~to~\SI{10}{\mega\hertz}~(Fig.~\ref{fig:GammaLamb}). The exceptionally broad tuning range makes it possible to accurately observe the Lamb shift, ranging from \SIrange{-8}{3}{\mega\hertz}. The tuning is controlled with a bias voltage, which shifts the relative chemical potential between the normal-metal and superconductor leads and activates the photon-assisted tunneling when the chemical potential is near the edge of the gap of the superconductor density of states~(Fig.~\ref{fig:teaser}). Finally, we verify our model by measuring the response of the coupling strength to changes in the normal-metal electron temperature~(Fig.~\ref{fig:T}).    
  
Figure~\ref{fig:teaser}a--b describes the measurement scheme~(Methods) and the samples, the fabrication of which is detailed in ref.~\onlinecite{Masuda16}. The resonator is capacitively coupled to a normal-metal island which is tunnel-coupled to two superconducting leads. An electron tunneling event between the island and the leads shifts the charge of the resonator by an amount of $\Delta Q=\alpha e$, where $\alpha\approx 1$ is a capacitance fraction defined in Fig.~\ref{fig:teaser} and $e$ is the elementary charge. A tunneling event couples different states of the resonator mode, and can lead to the creation and annihilation of photons. The rates of these processes are proportional to factors arising from the charge shift, junction transparency, and energy conservation~\cite{Silveri17prb} as detailed in Fig.~\ref{fig:teaser}c--d~(Methods). Note however that a linear resonator is not dephased by charge fluctuations. 

The resonator is probed through a \num{50}-\si{\ohm}~transmission line in a standard microwave reflection experiment~(Fig.~\ref{fig:teaser}a). The voltage reflection coefficient $\Gamma=|\Gamma|\ee^{-\ii \varphi}$ of a weak probe signal at the angular frequency $\omega_{\rm p}$ is given by
\begin{equation} 
  \label{eq:refl}
  \Gamma=\frac{\gamma_{\rm tr}-\gamma_{\rm T}-\gamma_0+ 2 \ii (\omega_{\rm p}-\omega_{\rm r})}{\gamma_{\rm tr}+\gamma_{\rm T}+\gamma_0-2 \ii (\omega_{\rm p}-\omega_{\rm r})}, 
\end{equation}
where $\gamma_{\rm tr}$ is the coupling strength to the transmission line and $\gamma_0$ represents the damping rate of the resonator by excess sources~(Methods). Figures~\ref{fig:GammaLamb}a--b show the magnitude of the measured reflection coefficient for Sample A and B (for the phase data see Supplementary Fig.~S3). At a given bias voltage, the minimum reflection occurring at $\omega_{\rm p}=\omega_{\rm r}$ yields the resonator frequency. The full width of the dip at half minimum equals the total coupling strength $\gamma_{\rm T}+\gamma_{\rm tr}+\gamma_0$, related to the loaded quality factor by $Q_\textrm{L}=\omega_{\rm r}/(\gamma_{\rm tr}+\gamma_{\rm T}+\gamma_0)$. At the critical points, where $\omega_{\rm p}=\omega_\textrm{r}$ and $\gamma_\textrm{T}+\gamma_0=\gamma_\textrm{tr}$ (black color in Fig.~\ref{fig:GammaLamb}a--b), the reflection ideally vanishes because of the impedance matching between the transmission line and the other electromagnetic environments of the resonator. Thus the full width of the dip $2\gamma_{\rm tr}$ gives accurately the coupling strength to the transmission line. The phase of the reflection coefficient exhibits a full $2\pi$ winding about the critical points~(Supplementary Fig.~S3). We extract the coupling strengths and the resonator frequency by fitting equation~\eqref{eq:refl} to the data (Methods).

\begin{table}
  \caption{\label{tab:parameters} Key device and model parameters. See Methods for details of the experimental determination of the parameters.}
  \begin{tabular}{l|c|c|c } 
    \hline
    Parameter & Symbol & Sample A & Sample B \\  
    \hline    
    \hline
    Resonator frequency (\si{\giga\hertz}) &  $\omega_{\rm r}/2\pi$ & $4.67$  & $8.54$  \\
    Charac. impedance (\si{\ohm}) & $Z_{\rm r}$ & $34.8$  & $34.8$  \\
    External coupling (\si{\mega\hertz}) &  $\gamma_{\rm tr}/2\pi$ & $3.7$  & $33.6$ \\ 
    Excess coupling (\si{\mega\hertz}) & $\gamma_0/2\pi$ &  $0.29$  & $10.6$ \\ 
    Coupling capacitance (\si{\femto\farad}) & $C_{\rm c}$ & $840$ & $780$  \\
    Island capacitance  (\si{\femto\farad}) &$C_{\rm \Sigma m}$ & $10$ & $10$  \\
    Superconductor gap (\si{\micro\elementarycharge\volt} ) & $\Delta $ & $215$ & $211$  \\
    Dynes parameter  &$\gamma_{\rm D}$ & \num[output-product=\times]{4xe-4} & \num[output-product=\times]{4xe-4}\\
    Junction conductance  (\si{\micro\siemens}) & $G_{\Sigma}$ & $71$ & $127$ \\
    Electron temperature (\si{\milli\kelvin}) & $T_{\rm N}$ & $170$ & $180$ \\
    \hline 
  \end{tabular}
\end{table}

Figures~\ref{fig:GammaLamb}c--d show the measured voltage-tunable coupling strength $\gamma_{\rm T}$ for the two samples. The characteristics of the coupling strength can be understood by considering tunneling at different bias voltages. If the junction is not biased and $\hbar\omega_{\rm r}\ll \Delta$, where the gap parameter $\Delta$ is defined in Fig.~\ref{fig:teaser}, the electron tunneling and the resulting coupling strength $\gamma_{\rm T}$ are suppressed by the small density of states in the superconductor gap~\cite{Dynes78}, quantified by the Dynes parameter $\gamma_{\rm D}\ll 1$. If the bias voltage is near the gap edge, the electron tunneling is efficiently assisted by thermal energy. As a result of thermal activation, the coupling strength $\gamma_{\rm T}$ increases exponentially as a function of the bias voltage, and reaches its maximum near the gap edge. At high bias voltages $|eV|/\Delta\gg 1$, the coupling strength $\gamma_{\rm T}$ saturates to the value $\alpha^2Z_{\rm r}G_{\Sigma}\omega_{\rm r}$, where $Z_{\rm r}$ is the characteristic impedance of the resonator and $G_{\Sigma}$ is the sum of the conductances of the two junctions~\cite{Silveri17prb}.  Consequently, we can tune the coupling strength $\gamma_{\rm T}$ by approximately three orders of magnitude with the bias voltage, which makes it possible to accurately measure the Lamb shift of the resonator. The measured values for the coupling strength are in good agreement with the theoretical model~\cite{Silveri17prb}~(Methods).

Figures~\ref{fig:GammaLamb}e--f show the observed shift of the resonator frequency $\omega_{\rm L}=\omega_{\rm r}-\omega_{\rm r}^{0}$ as a function of the bias voltage for the two samples. Here $\omega_{\rm r}^{0}$ is the resonator frequency at $V=0$. The natural frequency of a harmonic oscillator experiences a classical damping shift $\approx\gamma_{\rm T}^2/(8\omega_{\rm r})$ (not shown for clarity in figures) which, in our experimental setup,  is in the range of \SI{10}{\kilo\hertz} for Sample~A and \SI{100}{\kilo\hertz} for Sample~B and cannot explain the data. Interestingly the effective temperature of the environment increases as a function of the bias voltage~(see Supplementary Fig.~S4). However, contrary to the anharmonic systems, the harmonic oscillator has no ac Stark shift by the environment, that is, the energy level shifts are independent of the temperature of the environment~\cite{Carmichael}. Thus, we conclude that the observed shift of the resonator frequency is the Lamb shift induced by the broadband electromagnetic environment formed by the photon-assisted electron tunneling. In the following we confirm our conclusion by comparing the experimental results with a theoretical model.

We model the environment as a continuum of modes~\cite{Carmichael} characterized by their coupling strength $\gamma_{\rm T}(\omega)$  to the resonator, where $\omega$ refers to the frequency of a considered environmental mode. An environmental mode exchanges energy with the resonator only at resonance, being the principal mechanism for dissipation at the rate $\gamma_{\rm T}(\omega^0_{\rm r})$. Yet all the environmental modes are coupled to the system leading to the renormalization of its energy levels~\cite{Lamb47, Bethe47, Carmichael}. For a broadband environment, the corresponding dynamic Lamb shift for a harmonic oscillator is given by~\cite{Carmichael, Frisk14} 
\begin{equation} 
  \omega^{(\rm dyn)}_{\rm L}=-\textrm{PV}\int_0^\infty \frac{\Dn{\omega}}{2\pi}\left(\frac{\gamma_{\rm T}(\omega)}{\omega-\omega^{0}_{\rm r}}+\frac{\gamma_{\rm T}(\omega)}{\omega+\omega^{0}_{\rm r}}-2\frac{\gamma_{\rm T}(\omega)}{\omega}\right), \label{eq:Lamb}
\end{equation}
where $\textrm{PV}$ indicates the Cauchy principal value integration. The dynamic Lamb shift can be derived also from considering the broadband environment as a small electric admittance in parallel with the resonator and applying the Kramers--Kronig relations~\cite{LL_stat}~(see Methods and Supplementary Methods for details).

At bias values beyond the superconductor gap $eV/\Delta \gtrsim~2$, the electromagnetic environment formed by the photon-assisted tunneling at the normal-metal--insulator--superconductor junctions becomes ohmic~\cite{Silveri17prb}. Therefore, the coupling strength becomes linearly dependent on the frequency $\gamma_{\rm T}(\omega)= \alpha^2 Z_{\rm r}G_{\Sigma}\omega$. For an ohmic environment, the dynamic Lamb shift of a harmonic oscillator in equation~\eqref{eq:Lamb} vanishes~\cite{CaldeiraLeggett83}. In the experiments however, we study the frequency shifts with respect to the zero-voltage resonance, and hence the negative dynamic shift obtained from equation (2) at zero bias converts in experiments to a small positive shift  at high bias.

For Sample B, in addition to the dynamic shift we observe a shift that we identify as the static shift. We attribute this static shift to the effective elongation of the resonator mode caused by an increased current flow through the superconductor--insulator--normal-metal junction at high bias voltages. To the lowest order in the coupling strength, any static shift is given by $-\mu \gamma_{\rm T}/\pi$, where we obtain the proportionality constant $\mu=0.52$ for Sample~B. Due to the experimental experimental uncertainties, we cannot make a conclusive statement on the static shift in Sample~A. We attribute this effect to possible differences in the geometry and details of the junctions between the samples. As shown in Figs.~\ref{fig:GammaLamb}e--f this theory of the Lamb shift yields an excellent agreement with the measured data. Note that there are no free parameters in the theory curve of Fig.~\ref{fig:GammaLamb}e.

To further verify the applicability of the theoretical model of the photon-assisted tunneling, we study the response of the coupling strength~$\gamma_{\rm T}$ to the change in the normal-metal electron temperature~$T_{\rm N}$. We measure the critical bias point~$V_{\rm c}$, defined as the point at which $\gamma_{\rm T}+\gamma_0=\gamma_{\rm tr}$, where the reflection ideally vanishes. In elevated normal-metal electron temperatures, the thermally activated electron tunneling is enhanced, which leads to an increased coupling strength $\gamma_{\rm T}$ in the subgap~(Fig.~\ref{fig:T}a). As a result, the critical voltage moves to lower values~(Fig.~\ref{fig:T}b). In elevated temperatures, the density of quasiparticles is increased in the resonator, which leads to larger quasiparticle related losses~\cite{Gao08, Goetz16}. To account for this, the excess coupling strength $\gamma_0$ in Fig~\ref{fig:T} is assumed larger than in the low temperature data of Table~\ref{tab:parameters}. For simplicity, we assume it independent on temperature and voltage.  Overall, the good agreement between the measured and predicted critical voltages confirms that our model correctly captures the physics of the resonator environments. 

In conclusion, we observed the Lamb shift induced by a broadband environment in an engineered quantum system. The Lamb shift was observed to be tunable in regimes where both the dynamic and static parts significantly contribute. We demonstrated that the environmental coupling strength is tunable over more than three orders of magnitude, yet staying in the weak coupling regime between the system and the environment. To this end, we used bias-voltage-controlled electron tunneling in normal-metal--insulator--superconductor junctions, a device recently referred to as a quantum-circuit refrigerator~\cite{Tan16, Masuda16}. Our results are in excellent agreement with first-principles theory~\cite{Silveri17prb}, which verifies the validity of the model not only for the Lamb shift but also for the quantum-circuit refrigerator~\cite{Tan16, Masuda16}. Furthermore, our experiment expands the experimental operation regime of the quantum-circuit refrigerator to loaded quality factors up to $10^3$ and internal quality factors above \num{e4}, paving the way for rapid on-demand initialization of high-finesse quantum circuits, as well as the integration of the quantum-circuit refrigerator to synthetic quantum matter. With optimized sample parameters, our technique may allow us to systematically study the Lamb shift in the recently realized ultrastrong coupling regime~\cite{Forn-diaz16}.

\vspace*{0.15cm}

\textbf{Acknowledgments} We acknowledge discussions with Gianluigi Catelani, Aashish Clerk, Joonas Govenius, Hermann Grabert, and Jani Tuorila.  This research was financially supported by European Research Council under Grant No.~681311 (QUESS) and Marie Sk\l{}odowska-Curie Grant No.~795159; by Academy of Finland under its Centres of Excellence Program grants Nos.~312300, 312059 and grants Nos.~265675, 305237, 305306, 308161, 312300, 314302, 316551, 316619; JST ERATO Grant No.~JPMJER1601, JSPS KAKENHI Grant No.~18K03486 and by the Alfred Kordelin Foundation, the Emil Aaltonen Foundation, the Vilho, Yrj\"o and Kalle V\"ais\"al\"a Foundation, the Jane and Aatos Erkko Foundation, and the Technology Industries of Finland Centennial Foundation. We thank the provision of facilities and technical support by Aalto University at OtaNano~\---~Micronova Nanofabrication Centre.

\textbf{Author contributions} M.S. carried out the theoretical analysis and wrote the manuscript with input from all the authors. S.M., V.S., and M.J. conducted the experiments and analyzed the data. S.M. and K.Y.T. fabricated the samples. R.E.L., M.P., and J.G. contributed to the fabrication, development of the devices and the measurement scheme. L.G. fabricated the niobium layers. E.H., M.P., and J.G. contributed in the data analysis. E.H. and F.H. gave theory support. M.M. supervised the work in all respects. 

\textbf{Competing interests} The authors declare no competing interests. 

\textbf{Data availability} The data that support the findings of this study are available at \href{https://doi.org/10.5281/zenodo.1995361}{https://doi.org/10.5281/zenodo.1995361}.

\section*{Methods}

\textbf{Sample fabrication.} We fabricate the samples on \num{0.525}-\si{\milli\metre}-thick silicon wafers. The silicon surface is passivated by a \num{300}-\si{\nano\metre}-thick silicon oxide layer. We define the resonators by photolithography and ion etching of a \num{200}-\si{\nano\metre}-thick sputtered niobium layer, and then cover them  by a \num{50}-\si{\nano\metre}-thick layer of Al\textsubscript{2}O\textsubscript{3}. We produce the superconductor--insulator--normal-metal junctions with electron beam lithography followed by two-angle evaporation. More fabrication details can be found in ref.~\onlinecite{Masuda16}.\\

\textbf{Measurements.} We use a commercial dilution refrigerator to cool the samples down to the base temperature of \SI{10}{\milli\kelvin}. We attach the samples using vacuum grease to a sample holder with a printed circuit board, and bond them with aluminium wires. The printed circuit board is connected to the room-temperature setup by coaxial cables. The measurements are repeated multiple times.

The bias voltage is applied to the superconductor--insulator--normal-metal junctions by a battery-powered source. We measure the current through the junctions by a battery-powered transimpedance amplifier, which is connected to a voltmeter through an isolation amplifier.  We measure the reflection coefficient of the sample with a vector network analyser. Based on the power level of the vector network analyser and total attenuation, the power of the signal reaching the sample is around \SI{-100}{\deci\belm} (Supplementary Fig.~S2).   

The quasiparticle temperature of the superconducting leads and the electron temperature of the normal-metal island differ from the base temperature due to leakage through the radiation shields. They also depend on the level of the probe signal. However, no significant changes were noticed in the range of powers from \SIrange{-95}{-105}{\deci\belm}.

\textbf{Device and model parameters.} The resonator frequency~$\omega_{\textrm{r}}$, the  external coupling strength~$\gamma_{\textrm{tr}}$, the coupling strength~$\gamma_{\rm T}$, and the excess coupling strength~$\gamma_0$ are extracted from the reflection coefficient measurements using equation~\eqref{eq:refl} as follows. We assume that the measured reflection coefficient has a voltage-independent background arising, for example, from electrical delay or other reflections between the source and sample and between the sample and the vector network analyzer. To remove this background, we first divide a finite-voltage trace of the measured reflection coefficient by the zero-voltage trace. A trace means here a measurement of the reflection coefficient as a function of frequency by keeping the single-junction bias voltage fixed. The above-discussed division procedure yields us a normalized reflection coefficient illustrated in Supplementary Fig. S6a. Next, we fit to this result an equation of the form $r=\Gamma(V) / \Gamma(0)$, where $\Gamma$ is defined in equation~\eqref{eq:refl}  and $V$ is the voltage corresponding to the finite-voltage trace. However, the value of $V$ has no direct effect on the fit since we use the coupling strengths and the resonance frequencies as fitting parameters. We repeat this procedure for all $V$ used in the measured traces and obtain averaged parameter values for the zero-voltage reflection coefficient in equation~\eqref{eq:refl}, i.e., we obtain the background-subtracted trace $\Gamma'(0)$.  Subsequently we recalculate the background-corrected result for each measured finite-voltage traces as $\Gamma^{\prime}(V) = r\Gamma'(0)$ (Supplementary Fig. S6b).  This allows us to make a final fit of the data to equation~\eqref{eq:refl} at each bias voltage.  The results of this final fit are used in this manuscript.

  The error bars for the fits to equation~\eqref{eq:refl} are determined by drawing a circle of radius equal to root mean square fit error in the complex plane for the reflection coefficient. The center of the circle is placed at the resonance point of the least-squares fit according to equation~\eqref{eq:refl}. The confidence interval of each parameter is individually bounded by the condition that the resonance point  of a function following equation~\eqref{eq:refl} must lie within the circle when this parameter is varied but the other parameters correspond to the least-square fit. 

The capacitance of the normal-metal island to ground~$C_{\Sigma \textrm{m}}$ is a typical value for metallic islands with superconductor--insulator--normal-metal junctions~\cite{Tan16, Masuda16}. We calculate the impedance of the fundamental resonator mode as $Z_{\rm r} = (2/\pi) Z_0$. Here, $Z_0$ is the characteristic impedance of the coplanar waveguide structure obtained from the geometrical details of the device such as its center conductor and gap width. 

We extract the superconductor gap $\Delta$, the Dynes parameter $\gamma_{\textrm{D}}$, and the junction conductance $G_{\Sigma}$ from the current--voltage characteristics of the superconductor--insulator--normal-metal--insulator--superconductor junction~\cite{IngoldNazarov05}. The Dynes parameter $\gamma_{\textrm{D}}$ dominates the subgap current. The exact value of the junction conductance $G_{\Sigma}$ is obtained from the slope of the current--voltage curve at voltages beyond the superconductor gap and from the coupling strength at the high-bias values $\gamma_{\rm T}=\alpha^2 Z_{\rm r}G_{\Sigma}\omega_{\rm r}$. In refs.~\onlinecite{Tan16, Masuda16}, an extra pair of superconductor--insulator--normal-metal junctions served as a thermometer measuring the electron temperature of the normal-metal $T_{\rm N}$. From these measurements we estimate the electron temperature of the normal metal in the samples studied here. With a $10$-mK base temperature of the dilution refrigerator, the electron temperature $T_{\rm N}$ thermalizes to the values in the range from \SIrange{50}{200}{\milli\kelvin} for an unbiased junction. The exact value of the electron temperature $T_\textrm{N}$ in Table~\ref{tab:parameters} is obtained by the best fit of the theoretical result to the data in Figs.~\ref{fig:GammaLamb}c--d. For the higher cryostat temperatures in Fig.~\ref{fig:T}, we assumed that the electron temperature $T_\textrm{N}$ equals the base temperature. 

\textbf{Photon-assisted electron tunneling.} Reference~\onlinecite{Silveri17prb} details the theory of the photon-assisted tunneling at a normal-metal--insulator--superconductor junction. According to the theory, the photon-assisted tunneling forms an electromagnetic environment for a quantum circuit, such as a high-quality superconducting resonator. See also ref.~\onlinecite{IngoldNazarov05} for a general overview on tunneling at nanostructures. For completeness, we present here the main results of the theory,  namely the coupling strength and the effective temperature of the electromagnetic environment. Importantly, we extend here the theory by the derivation of the Lamb shift. 

We consider a normal-metal--insulator--superconductor junction at the energy bias $E$ and define a rate function
\begin{equation}
\label{eq:PEforw}
\normrateright(E) = \frac{1}{h}\int \Dn{\varepsilon}\, n_{\rm S}(\varepsilon) [1-f_{\rm S}(\varepsilon)]f_{\nmd}(\varepsilon-E),
\end{equation}
where $\varepsilon$ denotes electron energy. The function $\normrateright(E)$ gives the normalized rate of forward quasiparticle tunneling for a junction with conductance $G$ equal to half of the conductance quantum $G_0=2e^2/h$. The tunneling rate is dictated by the occupations of the normal-metal and superconductors through the Fermi functions, $f_{\rm N}(\varepsilon)$ and $f_{\rm S}(\varepsilon)$, respectively, as well as by the normalized quasiparticle density of the states in the superconductor
\begin{equation}
n_{\rm S}(\varepsilon)=\left\vert\textrm{Re}\left\{\frac{\varepsilon+i\gamma_{\rm D}\Delta}{\sqrt{(\varepsilon+i\gamma_{\rm D}\Delta)^2-\Delta^2}}\right\}\right\vert, \label{eq:dos}
\end{equation}
where $\Delta$ is the superconductor gap parameter and $\gamma_{\rm D}$ is the Dynes parameter~(Table~\ref{tab:parameters}) characterizing the subgap density of states $n_{\rm S}(0)\approx \gamma_{\rm D}$. A photon-assisted tunneling event shifts the charge of the resonator by an amount of $\Delta Q=\alpha e$, where $\alpha= C_{\rm c}/(C_{\rm c}+C_{\rm \Sigma m})$ is a capacitance fraction of the normal-metal island. The charge shift induces transitions from the resonator energy eigenstate $\ket{m}$ to the eigenstate $\ket{m^\prime}$ ($\ell=m^\prime-m \ge 0 $) through the matrix element
\begin{align}
|M_{m m^\prime}|^2 &=\left|\int\psi_{m^\prime}^\ast \left(Q-\alpha e\right) \psi_m\left(Q\right)  \Dn{Q}\right|^2 \notag\\& = \ee^{-\rho} \rho^\ell \frac{m^\prime !}{m!} \left[L^{\ell}_{m^\prime }(\rho)\right]^2 
\end{align}
where $\psi_m(Q)=\braket{Q|m}$ are the resonator energy eigenstates represented in the charge basis, $\rho=\pi \alpha^2 Z_{\rm r}/R_{\rm K}$ is an interaction parameter expressed in terms of the characteristic impedance $Z_{\rm r}$ of the resonator, and $L^{\ell}_{m^\prime }(\rho)$ denote the generalized Laguerre polynomials.

The resonator transition rate becomes~\cite{Silveri17prb} 
\begin{equation}
  \Gamma_{m,m^\prime}(V) = |M_{m m^\prime}|^2 R_{\rm K} G_{\Sigma} \sum_{\tau=\pm 1 } \normrateright \left(\tau eV+\hbar \omega_{\rm r} \ell-E_{\nmd} \right),
\end{equation}
where we have assumed that two superconductor--insulator--normal-metal junctions of the superconductor--insulator--normal-metal--insulator--superconductor construction are sufficiently identical, the electrodes are at equal temperatures, and that the charging energy $E_{\rm N}=e^2/2(C_{\rm c}+C_{\rm \Sigma m}) \sim h \times$ \SI{10}{\mega\hertz} of the normal-metal island is the smallest of the relevant energy scales of the setup ($\Delta$, $\hbar \omega_{\rm r}$, and $k_{\rm B}T_{\rm N}$). In a typical experimental scenario, the interaction parameter $\rho$ is well below unity since $Z_{\rm r}\ll R_{\rm K}$. Thus, at low powers the dominant transitions are those between adjacent states $\Gamma_{m, m-1}$ and $\Gamma_{m, m+1}$. In this case, we characterize the electromagnetic environment through its coupling strength $\gamma_{\rm T}$ 
\begin{equation}
  \gamma_{\rm T}(V, \omega_{\rm r}) = \pi \alpha^2 Z_{\rm r} G_{\Sigma} \sum_{\ell,\tau=\pm 1} \ell\normrateright \left(\tau eV+\ell\hbar \omega_{\rm r}-E_{\nmd} \right),  \label{eq:gammaa}\\
\end{equation}
as well as the effective mode temperature $T_{\rm T}$,
\begin{equation}
T_{\rm T}(V, \omega_{\rm r}) = \frac{\hbar\omega_{\rm r}}{k_{\rm B}}\left[ \ln \left( \frac{ \sum_{\tau=\pm 1}  \normrateright \left(\tau eV+\hbar \omega_{\rm r}-E_{\nmd} \right)}{\sum_{\tau =\pm 1}  \normrateright \left(\tau eV-\hbar \omega_{\rm r}-E_{\nmd} \right)}\right) \right]^{-1},  \label{eq:T}
\end{equation}
which are defined through the mapping $\Gamma_{m, m-1}  =\gamma_{\rm T}(N_{\rm T}+1)m$ and $\Gamma_{m, m+1}  =\gamma_{\rm T} N_{\rm T} (m+1)$ of the transition rates, where the mean number of excitations $N_{\rm T}=1/[\ee^{\hbar \omega_{\rm r}/(k_{\rm B}T_{\rm T})}-1]$ defines the effective mode temperature $T_{\rm T}$. Here, $k_{\rm B}$ is the Boltzmann constant.

The quasiparticle tunneling across the normal-metal--insulator--superconductor junction, characterized by the tunneling rate function ${\normrateright}(E)$ in equation~\eqref{eq:PEforw}, defines the dependence of the coupling strength $\gamma_{\rm T}$ on the resonator frequency  $\omega_{\rm r}$ and the bias voltage $eV$. To completely map the tunneling rate function, one needs to measure both the coupling strength $\gamma_{\rm T}$ and $T_{\rm T}$.  In refs.~\onlinecite{Tan16, Masuda16, Silveri17prb}, we have probed these quantities with excellent agreement with the theoretical equations~\eqref{eq:gammaa} and~\eqref{eq:T}. Here, we probe the dependence of the coupling strength $\gamma_{\rm T}$ on the bias voltage (Fig.~\ref{fig:GammaLamb}c--d)  and observe that the experimental results are in accordance with the theory.  Combining these observations, we verify that our model is valid. The dependence of the coupling strength on the resonator frequency $\omega_{\rm r}$ originates from the same rate function summing to the voltage, thus validating the use of equation~\eqref{eq:gammaa} in the calculation of the Lamb shift  where it is used for a broad range of frequencies.
 
By applying second-order time-independent perturbation theory, we can also derive the Lamb shift of the resonator caused by the quasiparticle tunneling through the normal-metal--insulator--superconductor junctions. The derivation, detailed in Supplementary Methods 1, follows the assumptions and guidelines of refs.~\onlinecite{IngoldNazarov05, Silveri17prb}, summarized above. The resulting Lamb shift is given by
\begin{align} 
\omega_{\rm{L}}(V, \omega_{\rm r}) =  - \textrm{PV}\int_{0}^\infty \frac{\Dn{\omega}}{2\pi}\bigg[\frac{\gamma_{\rm T}(V, \omega)}{\omega-\omega_{\rm r}} &+\frac{\gamma_{\rm T}(V, \omega)}{\omega+\omega_{\rm r}}\notag \\ &-\frac{2\gamma_{\rm T}(V, \omega)}{\omega}\bigg].
\end{align}
The two first terms originate from the photon-assisted tunneling processes. Hence, they depend on the resonator frequency. The third term originates from the elastic tunneling and is independent of the resonator frequency. Importantly, the elastic tunneling affects the energy levels despite of exchanging no energy with the resonator and having no contribution on the coupling strength~$\gamma_{\rm T}$ or the effective temperature~$T_{\rm T}$. 


\begin{thebibliography}{10}
\expandafter\ifx\csname url\endcsname\relax
  \def\url#1{\texttt{#1}}\fi
\expandafter\ifx\csname urlprefix\endcsname\relax\def\urlprefix{URL }\fi
\providecommand{\bibinfo}[2]{#2}
\providecommand{\eprint}[2][]{\url{#2}}

\bibitem{Lamb47}
\bibinfo{author}{Lamb, W.~E.} \& \bibinfo{author}{Retherford, R.~C.}
\newblock \bibinfo{title}{Fine {structure} of the {hydrogen} {atom} by a
  {microwave} {method}}.
\newblock \emph{\bibinfo{journal}{Phys. Rev.}} \textbf{\bibinfo{volume}{72}},
  \bibinfo{pages}{241--243} (\bibinfo{year}{1947}).
\newblock \urlprefix\url{https://link.aps.org/doi/10.1103/PhysRev.72.241}.

\bibitem{Bethe47}
\bibinfo{author}{Bethe, H.~A.}
\newblock \bibinfo{title}{The electromagnetic shift of energy levels}.
\newblock \emph{\bibinfo{journal}{Phys. Rev.}} \textbf{\bibinfo{volume}{72}},
  \bibinfo{pages}{339} (\bibinfo{year}{1947}).

\bibitem{Heinzen87}
\bibinfo{author}{Heinzen, D.~J.} \& \bibinfo{author}{Feld, M.~S.}
\newblock \bibinfo{title}{Vacuum {radiative} {level} {shift} and
  {spontaneous}-{emission} {linewidth} of an {atom} in an {optical}
  {resonator}}.
\newblock \emph{\bibinfo{journal}{Phys. Rev. Lett.}}
  \textbf{\bibinfo{volume}{59}}, \bibinfo{pages}{2623--2626}
  (\bibinfo{year}{1987}).
\newblock \urlprefix\url{https://link.aps.org/doi/10.1103/PhysRevLett.59.2623}.

\bibitem{Brune94}
\bibinfo{author}{Brune, M.} \emph{et~al.}
\newblock \bibinfo{title}{From {Lamb} shift to light shifts: {Vacuum} and
  subphoton cavity fields measured by atomic phase sensitive detection}.
\newblock \emph{\bibinfo{journal}{Phys. Rev. Lett.}}
  \textbf{\bibinfo{volume}{72}}, \bibinfo{pages}{3339--3342}
  (\bibinfo{year}{1994}).
\newblock \urlprefix\url{https://link.aps.org/doi/10.1103/PhysRevLett.72.3339}.

\bibitem{Marrocco98}
\bibinfo{author}{Marrocco, M.}, \bibinfo{author}{Weidinger, M.},
  \bibinfo{author}{Sang, R.~T.} \& \bibinfo{author}{Walther, H.}
\newblock \bibinfo{title}{Quantum {electrodynamic} {shifts} of {Rydberg}
  {energy} {levels} between {parallel} {metal} {plates}}.
\newblock \emph{\bibinfo{journal}{Phys. Rev. Lett.}}
  \textbf{\bibinfo{volume}{81}}, \bibinfo{pages}{5784--5787}
  (\bibinfo{year}{1998}).
\newblock \urlprefix\url{https://link.aps.org/doi/10.1103/PhysRevLett.81.5784}.

\bibitem{Carmichael}
\bibinfo{author}{Carmichael, H.~J.}
\newblock \emph{\bibinfo{title}{Statistical {methods} in {quantum} {optics}~1}}
  (\bibinfo{publisher}{Springer}, \bibinfo{address}{Berlin, Heidelberg},
  \bibinfo{year}{1999}).
\newblock \urlprefix\url{http://link.springer.com/10.1007/978-3-662-03875-8}.

\bibitem{Gisin07}
\bibinfo{author}{Gisin, N.} \& \bibinfo{author}{Thew, R.}
\newblock \bibinfo{title}{Quantum communication}.
\newblock \emph{\bibinfo{journal}{Nat. Photonics}}
  \textbf{\bibinfo{volume}{1}}, \bibinfo{pages}{165--171}
  (\bibinfo{year}{2007}).

\bibitem{Ladd10}
\bibinfo{author}{Ladd, T.~D.} \emph{et~al.}
\newblock \bibinfo{title}{Quantum computers}.
\newblock \emph{\bibinfo{journal}{Nature}} \textbf{\bibinfo{volume}{464}},
  \bibinfo{pages}{45--53} (\bibinfo{year}{2010}).
\newblock
  \urlprefix\url{http://www.nature.com/nature/journal/v464/n7285/full/nature08812.html}.

\bibitem{Rentrop16}
\bibinfo{author}{Rentrop, T.} \emph{et~al.}
\newblock \bibinfo{title}{Observation of the {phononic} {Lamb} {shift} with a
  {synthetic} {vacuum}}.
\newblock \emph{\bibinfo{journal}{Phys. Rev. X}} \textbf{\bibinfo{volume}{6}},
  \bibinfo{pages}{041041} (\bibinfo{year}{2016}).
\newblock \urlprefix\url{https://link.aps.org/doi/10.1103/PhysRevX.6.041041}.

\bibitem{Fragner08}
\bibinfo{author}{Fragner, A.} \emph{et~al.}
\newblock \bibinfo{title}{Resolving {vacuum} {fluctuations} in an {electrical}
  {circuit} by {measuring} the {Lamb} {shift}}.
\newblock \emph{\bibinfo{journal}{Science}} \textbf{\bibinfo{volume}{322}},
  \bibinfo{pages}{1357--1360} (\bibinfo{year}{2008}).
\newblock \urlprefix\url{http://science.sciencemag.org/content/322/5906/1357}.

\bibitem{Yoshihara18}
\bibinfo{author}{Yoshihara, F.} \emph{et~al.}
\newblock \bibinfo{title}{Inversion of qubit energy levels in qubit-oscillator
  circuits in the deep-strong-coupling regime}.
\newblock \emph{\bibinfo{journal}{Phys. Rev. Lett.}}
  \textbf{\bibinfo{volume}{120}}, \bibinfo{pages}{183601}
  (\bibinfo{year}{2018}).

\bibitem{Mirhosseini18}
\bibinfo{author}{Mirhosseini, M.} \emph{et~al.}
\newblock \bibinfo{title}{Superconducting metamaterials for waveguide quantum
  electrodynamics}.
\newblock \emph{\bibinfo{journal}{Nat. Commun.}} \textbf{\bibinfo{volume}{9}},
  \bibinfo{pages}{3706} (\bibinfo{year}{2018}).
\newblock \urlprefix\url{https://www.nature.com/articles/s41467-018-06142-z}.

\bibitem{Partanen16}
\bibinfo{author}{Partanen, M.} \emph{et~al.}
\newblock \bibinfo{title}{Quantum-limited heat conduction over macroscopic
  distances}.
\newblock \emph{\bibinfo{journal}{Nat. Phys.}} \textbf{\bibinfo{volume}{12}},
  \bibinfo{pages}{460--464} (\bibinfo{year}{2016}).
\newblock \urlprefix\url{https://www.nature.com/articles/nphys3642}.

\bibitem{Tan16}
\bibinfo{author}{Tan, K.~Y.} \emph{et~al.}
\newblock \bibinfo{title}{Quantum-circuit refrigerator}.
\newblock \emph{\bibinfo{journal}{Nat. Commun.}} \textbf{\bibinfo{volume}{8}},
  \bibinfo{pages}{15189} (\bibinfo{year}{2017}).

\bibitem{Masuda16}
\bibinfo{author}{Masuda, S.} \emph{et~al.}
\newblock \bibinfo{title}{Observation of microwave absorption and emission from
  incoherent electron tunneling through a normal-metal-insulator-superconductor
  junction}.
\newblock \emph{\bibinfo{journal}{Sci. Rep.}} \textbf{\bibinfo{volume}{8}},
  \bibinfo{pages}{3966} (\bibinfo{year}{2018}).

\bibitem{Houck12}
\bibinfo{author}{Houck, A.~A.}, \bibinfo{author}{T{\"u}reci, H.~E.} \&
  \bibinfo{author}{Koch, J.}
\newblock \bibinfo{title}{On-chip quantum simulation with superconducting
  circuits}.
\newblock \emph{\bibinfo{journal}{Nat. Phys.}} \textbf{\bibinfo{volume}{8}},
  \bibinfo{pages}{292--299} (\bibinfo{year}{2012}).
\newblock
  \urlprefix\url{http://www.nature.com/nphys/journal/v8/n4/abs/nphys2251.html}.

\bibitem{Fitzpatrick17}
\bibinfo{author}{Fitzpatrick, M.}, \bibinfo{author}{Sundaresan, N.~M.},
  \bibinfo{author}{Li, A.~C.}, \bibinfo{author}{Koch, J.} \&
  \bibinfo{author}{Houck, A.~A.}
\newblock \bibinfo{title}{Observation of a {dissipative} {phase} {transition}
  in a {one}-{dimensional} {circuit} {QED} {lattice}}.
\newblock \emph{\bibinfo{journal}{Phys. Rev. X}} \textbf{\bibinfo{volume}{7}},
  \bibinfo{pages}{011016} (\bibinfo{year}{2017}).
\newblock \urlprefix\url{https://link.aps.org/doi/10.1103/PhysRevX.7.011016}.

\bibitem{Ma18}
\bibinfo{author}{Ma, R.} \emph{et~al.}
\newblock \bibinfo{title}{A {dissipatively} {stabilized} {Mott} {insulator} of
  {photons}}.
\newblock \emph{\bibinfo{journal}{Preprint at http://arxiv.org/abs/1807.11342}}
   (\bibinfo{year}{2018}).
\newblock \urlprefix\url{http://arxiv.org/abs/1807.11342}.

\bibitem{Silveri17prb}
\bibinfo{author}{Silveri, M.}, \bibinfo{author}{Grabert, H.},
  \bibinfo{author}{Masuda, S.}, \bibinfo{author}{Tan, K.~Y.} \&
  \bibinfo{author}{M\"ott\"onen, M.}
\newblock \bibinfo{title}{Theory of quantum-circuit refrigeration by
  photon-assisted electron tunneling}.
\newblock \emph{\bibinfo{journal}{Phys. Rev. B}} \textbf{\bibinfo{volume}{96}},
  \bibinfo{pages}{094524} (\bibinfo{year}{2017}).
\newblock \urlprefix\url{https://link.aps.org/doi/10.1103/PhysRevB.96.094524}.

\bibitem{Gramich2011}
\bibinfo{author}{Gramich, V.}, \bibinfo{author}{Solinas, P.},
  \bibinfo{author}{M\"ott\"onen, M.}, \bibinfo{author}{Pekola, J.~P.} \&
  \bibinfo{author}{Ankerhold, J.}
\newblock \bibinfo{title}{Measurement scheme for the {Lamb} shift in a
  superconducting circuit with broadband environment}.
\newblock \emph{\bibinfo{journal}{Phys. Rev. A}} \textbf{\bibinfo{volume}{84}},
  \bibinfo{pages}{052103} (\bibinfo{year}{2011}).
\newblock \urlprefix\url{https://link.aps.org/doi/10.1103/PhysRevA.84.052103}.

\bibitem{Paraoanu06}
\bibinfo{author}{Paraoanu, G.~S.}
\newblock \bibinfo{title}{Microwave-induced coupling of superconducting
  qubits}.
\newblock \emph{\bibinfo{journal}{Phys. Rev. B}} \textbf{\bibinfo{volume}{74}},
  \bibinfo{pages}{140504} (\bibinfo{year}{2006}).
\newblock \urlprefix\url{https://link.aps.org/doi/10.1103/PhysRevB.74.140504}.

\bibitem{Rigetti10}
\bibinfo{author}{Rigetti, C.} \& \bibinfo{author}{Devoret, M.}
\newblock \bibinfo{title}{Fully microwave-tunable universal gates in
  superconducting qubits with linear couplings and fixed transition
  frequencies}.
\newblock \emph{\bibinfo{journal}{Phys. Rev. B}} \textbf{\bibinfo{volume}{81}},
  \bibinfo{pages}{134507} (\bibinfo{year}{2010}).
\newblock \urlprefix\url{https://link.aps.org/doi/10.1103/PhysRevB.81.134507}.

\bibitem{Kerckhoff10}
\bibinfo{author}{Kerckhoff, J.}, \bibinfo{author}{Nurdin, H.~I.},
  \bibinfo{author}{Pavlichin, D.~S.} \& \bibinfo{author}{Mabuchi, H.}
\newblock \bibinfo{title}{Designing {quantum} {memories} with {embedded}
  {control}: {Photonic} {circuits} for {autonomous} {quantum} {error}
  {correction}}.
\newblock \emph{\bibinfo{journal}{Phys. Rev. Lett.}}
  \textbf{\bibinfo{volume}{105}}, \bibinfo{pages}{040502}
  (\bibinfo{year}{2010}).
\newblock
  \urlprefix\url{http://link.aps.org/doi/10.1103/PhysRevLett.105.040502}.

\bibitem{Kapit15}
\bibinfo{author}{Kapit, E.}, \bibinfo{author}{Chalker, J.~T.} \&
  \bibinfo{author}{Simon, S.~H.}
\newblock \bibinfo{title}{Passive correction of quantum logical errors in a
  driven, dissipative system: {A} blueprint for an analog quantum code fabric}.
\newblock \emph{\bibinfo{journal}{Phys. Rev. A}} \textbf{\bibinfo{volume}{91}},
  \bibinfo{pages}{062324} (\bibinfo{year}{2015}).
\newblock \urlprefix\url{https://link.aps.org/doi/10.1103/PhysRevA.91.062324}.

\bibitem{Geerlings13}
\bibinfo{author}{Geerlings, K.} \emph{et~al.}
\newblock \bibinfo{title}{Demonstrating a {driven} {reset} {protocol} for a
  {superconducting} {qubit}}.
\newblock \emph{\bibinfo{journal}{Phys. Rev. Lett.}}
  \textbf{\bibinfo{volume}{110}}, \bibinfo{pages}{120501}
  (\bibinfo{year}{2013}).
\newblock
  \urlprefix\url{https://link.aps.org/doi/10.1103/PhysRevLett.110.120501}.

\bibitem{Partanen18}
\bibinfo{author}{Partanen, M.} \emph{et~al.}
\newblock \bibinfo{title}{Optimized heat transfer at exceptional points in
  quantum circuits}.
\newblock \emph{\bibinfo{journal}{Preprint at
  https://arxiv.org/abs/1812.02683}}  (\bibinfo{year}{2018}).
\newblock \urlprefix\url{https://arxiv.org/abs/1812.02683}.

\bibitem{Weiss}
\bibinfo{author}{Weiss, U.}
\newblock \emph{\bibinfo{title}{Quantum dissipative systems}}
  (\bibinfo{publisher}{World Scientific}, \bibinfo{address}{Berlin,
  Heidelberg}, \bibinfo{year}{2012}).

\bibitem{Frisk14}
\bibinfo{author}{Frisk~Kockum, A.}, \bibinfo{author}{Delsing, P.} \&
  \bibinfo{author}{Johansson, G.}
\newblock \bibinfo{title}{Designing frequency-dependent relaxation rates and
  {Lamb} shifts for a giant artificial atom}.
\newblock \emph{\bibinfo{journal}{Phys. Rev. A}} \textbf{\bibinfo{volume}{90}},
  \bibinfo{pages}{013837} (\bibinfo{year}{2014}).
\newblock \urlprefix\url{https://link.aps.org/doi/10.1103/PhysRevA.90.013837}.

\bibitem{IngoldNazarov05}
\bibinfo{author}{Ingold, G.-L.} \& \bibinfo{author}{Nazarov, Y.~V.}
\newblock \bibinfo{title}{Charge tunneling rates in ultrasmall junctions}.
\newblock In \bibinfo{editor}{Grabert, H.} \& \bibinfo{editor}{Devoret, M.~H.}
  (eds.) \emph{\bibinfo{booktitle}{Single charge tunneling: Coulomb blockade
  phenomena in nanostructures}} (\bibinfo{publisher}{Plenum},
  \bibinfo{address}{New York}, \bibinfo{year}{1992}).
\newblock \urlprefix\url{http://arxiv.org/abs/cond-mat/0508728}.

\bibitem{Dynes78}
\bibinfo{author}{Dynes, R.~C.}, \bibinfo{author}{Narayanamurti, V.} \&
  \bibinfo{author}{Garno, J.~P.}
\newblock \bibinfo{title}{Direct measurement of quasiparticle-lifetime
  broadening in a strong-coupled superconductor}.
\newblock \emph{\bibinfo{journal}{Phys. Rev. Lett.}}
  \textbf{\bibinfo{volume}{41}}, \bibinfo{pages}{1509} (\bibinfo{year}{1978}).

\bibitem{LL_stat}
\bibinfo{author}{Landau, L.~D.} \& \bibinfo{author}{Lifshitz, E.~M.}
\newblock \emph{\bibinfo{title}{Statistical Physics, Part~1}}
  (\bibinfo{publisher}{Pergamon}, \bibinfo{address}{Oxford},
  \bibinfo{year}{1980}).

\bibitem{CaldeiraLeggett83}
\bibinfo{author}{Caldeira, A.~O.} \& \bibinfo{author}{Leggett, A.~J.}
\newblock \bibinfo{title}{Quantum tunnelling in a dissipative system}.
\newblock \emph{\bibinfo{journal}{Ann. Phys.}} \textbf{\bibinfo{volume}{149}},
  \bibinfo{pages}{374} (\bibinfo{year}{1983}).

\bibitem{Gao08}
\bibinfo{author}{Gao, J.} \emph{et~al.}
\newblock \bibinfo{title}{Equivalence of the {effects} on the {complex}
  {conductivity} of {superconductor} due to {temperature} {change} and
  {external} {pair} {breaking}}.
\newblock \emph{\bibinfo{journal}{J. Low Temp. Phys.}}
  \textbf{\bibinfo{volume}{151}}, \bibinfo{pages}{557--563}
  (\bibinfo{year}{2008}).
\newblock \urlprefix\url{http://link.springer.com/10.1007/s10909-007-9688-z}.

\bibitem{Goetz16}
\bibinfo{author}{Goetz, J.} \emph{et~al.}
\newblock \bibinfo{title}{Loss mechanisms in superconducting thin film
  microwave resonators}.
\newblock \emph{\bibinfo{journal}{J. App. Phys.}}
  \textbf{\bibinfo{volume}{119}}, \bibinfo{pages}{015304}
  (\bibinfo{year}{2016}).
\newblock \urlprefix\url{http://doi.org/10.1063/1.4939299}.

\bibitem{Forn-diaz16}
\bibinfo{author}{Forn-D\'iaz, P.} \emph{et~al.}
\newblock \bibinfo{title}{Ultrastrong coupling of a single artificial atom to
  an electromagnetic continuum in the nonperturbative regime}.
\newblock \emph{\bibinfo{journal}{Nat. Phys.}} \textbf{\bibinfo{volume}{13}},
  \bibinfo{pages}{39} (\bibinfo{year}{2017}).
\newblock \urlprefix\url{http://doi.org/10.1038/nphys3905}.

\end{thebibliography}
\end{document}


\title{Supplementary information: \\ \vspace*{6pt} Observation of a broadband Lamb shift in an engineered quantum system} 
\author{Matti Silveri}
\author{Shumpei Masuda}
\author{Vasilii Sevriuk}
\author{Kuan Y. Tan}
\author{M\'at\'e Jenei}
\author{Eric Hyypp\"a} 
\author{Fabian~Hassler}
\author{Matti~Partanen}
\author{Jan Goetz}
\author{Russell E. Lake}
\author{Leif Gr\"onberg}
\author{Mikko M\"{o}tt\"{o}nen}

\date{\today}
\maketitle

\begin{figure}[h]  
  \centering  
  \includegraphics[height=0.31\linewidth]{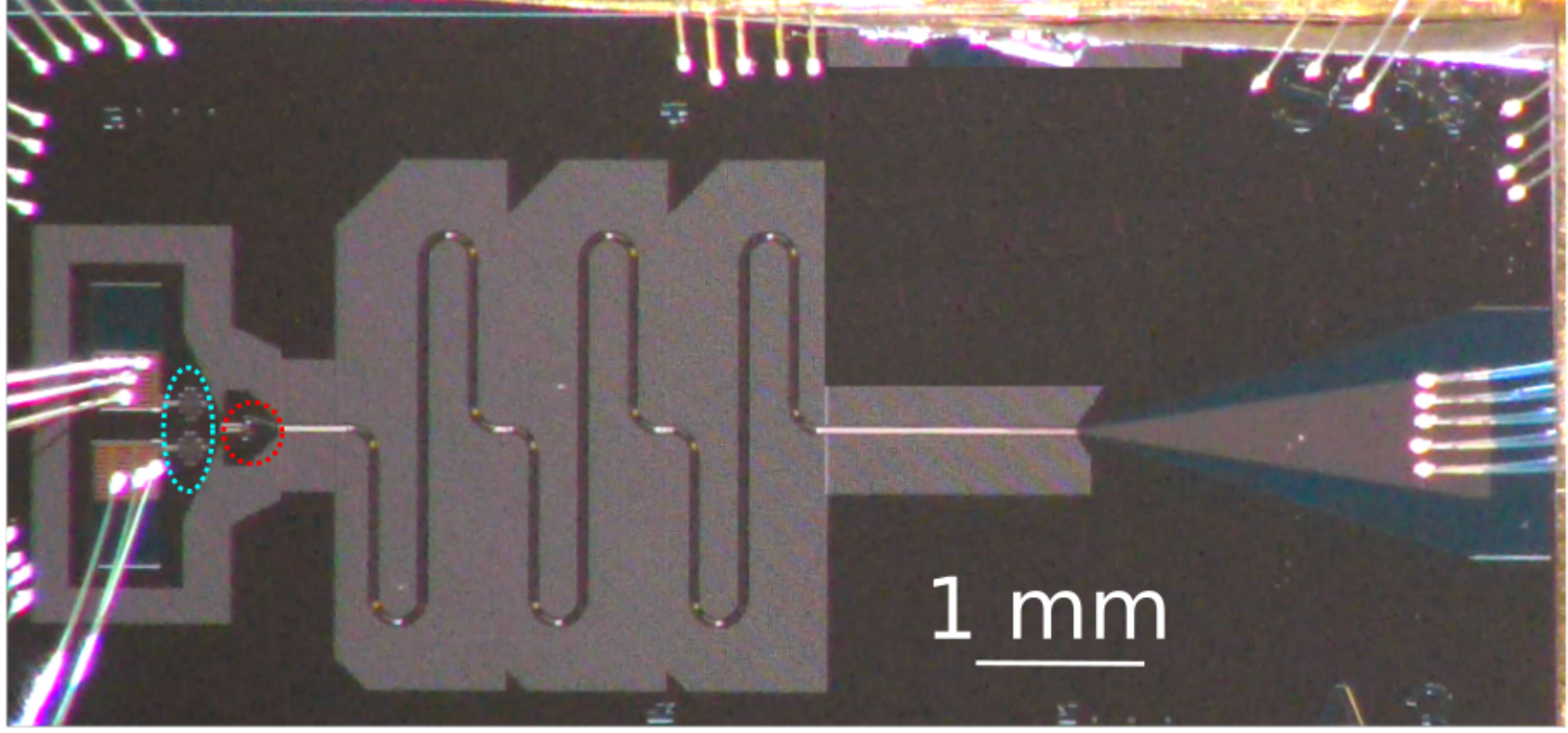}
  \caption{\textbf{Sample A}. Image of the device showing from left to right the $RC$ filters (cyan circle), the quantum-circuit refrigerator (red circle) comprising of two superconductor--insulator--normal-metal tunnel junctions, the coplanar-waveguide resonator (meandering line), and the transmission line  (horizontal line with tapering). Sample~B is similar to Sample~A except that it lacks the $RC$ filters. }
\end{figure}
\newpage  

\begin{figure}[h]  
  \centering
  \includegraphics[height=0.5\linewidth]{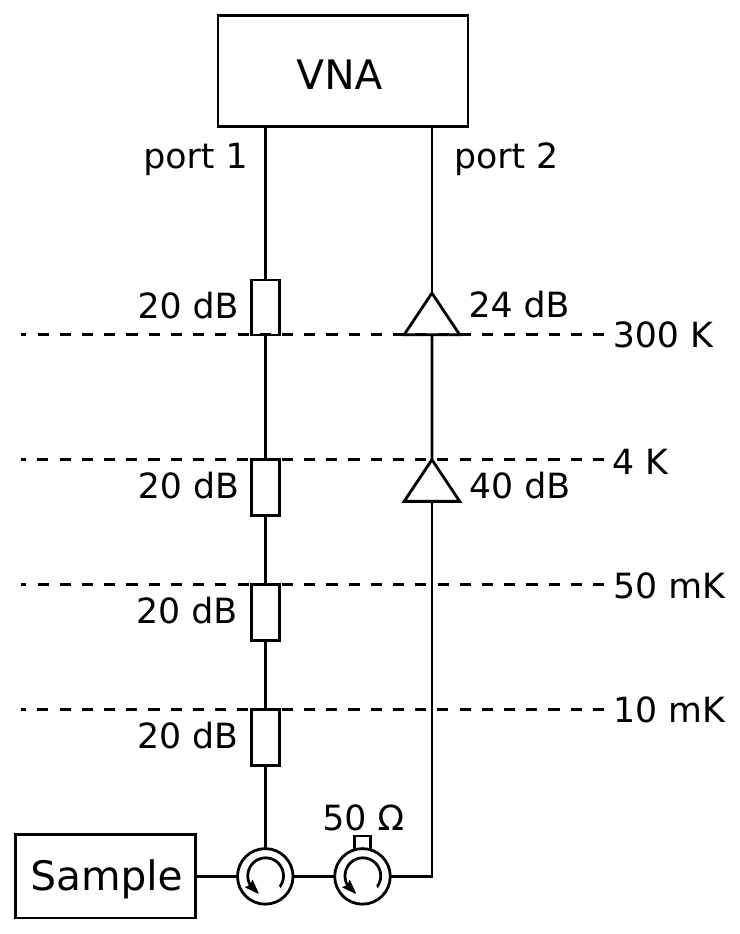}
  \caption{\textbf{Measurement setup for Sample B}. The signal going from port 1 of the vector network analyzer (VNA) to the sample is attenuated at different temperature stages as indicated, and the signal going from the sample to port 2 of  the VNA is amplified at \SI{4}{\kelvin} and at room temperature. A circulator is used to guide the input signal to the sample and the reflected signal via the amplifiers back to the VNA. Another circulator with a \num{50}-\si{\ohm} termination isolates the sample from the noise coming from the amplifier. The setup for Sample A is similar. }
\end{figure}
\newpage

\begin{figure}[h]  
  \centering
  \includegraphics[height=0.35\linewidth]{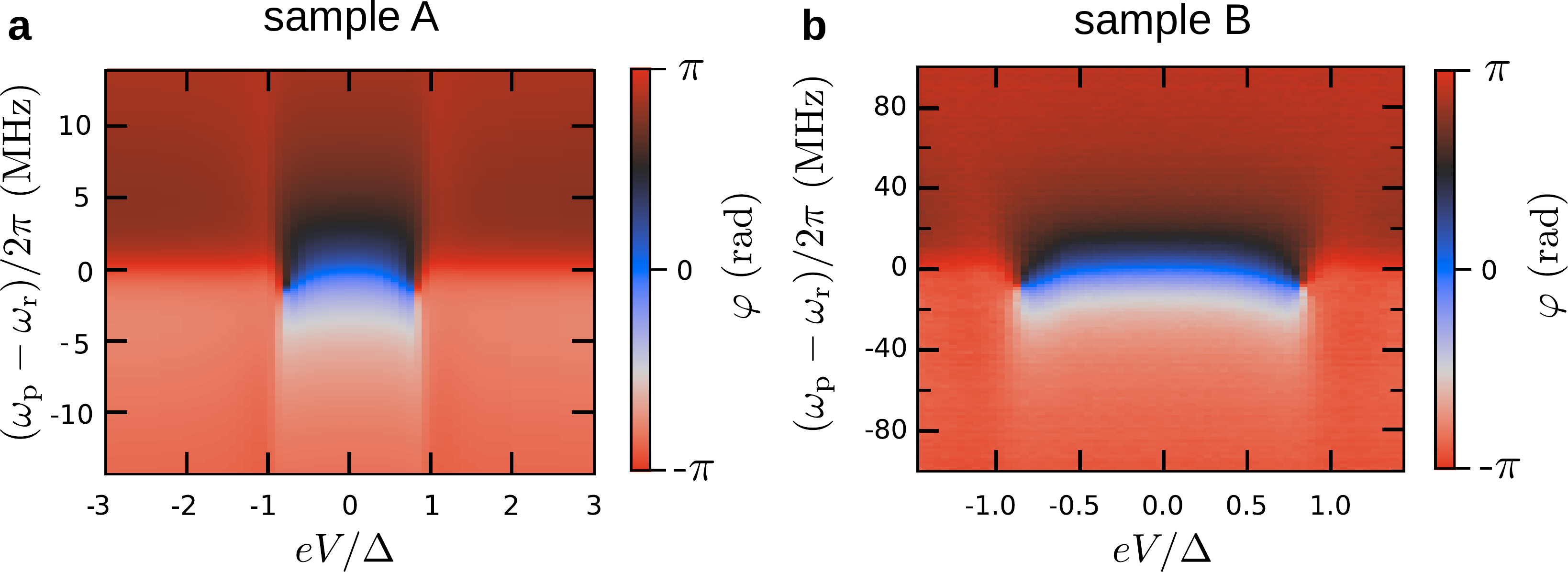}
  \caption{\textbf{Voltage reflection coefficient}. \textbf{a, b,} Phase $\varphi$ of the voltage reflection coefficient $\Gamma=|\Gamma|\exp(-\ii \varphi)$ as a function of the probe frequency $\omega_{\rm p}$ and the single-junction bias voltage $V$ for Samples~A and~B.}
\end{figure}
\newpage

\begin{figure}
  \centering
  \includegraphics[height=0.35\linewidth]{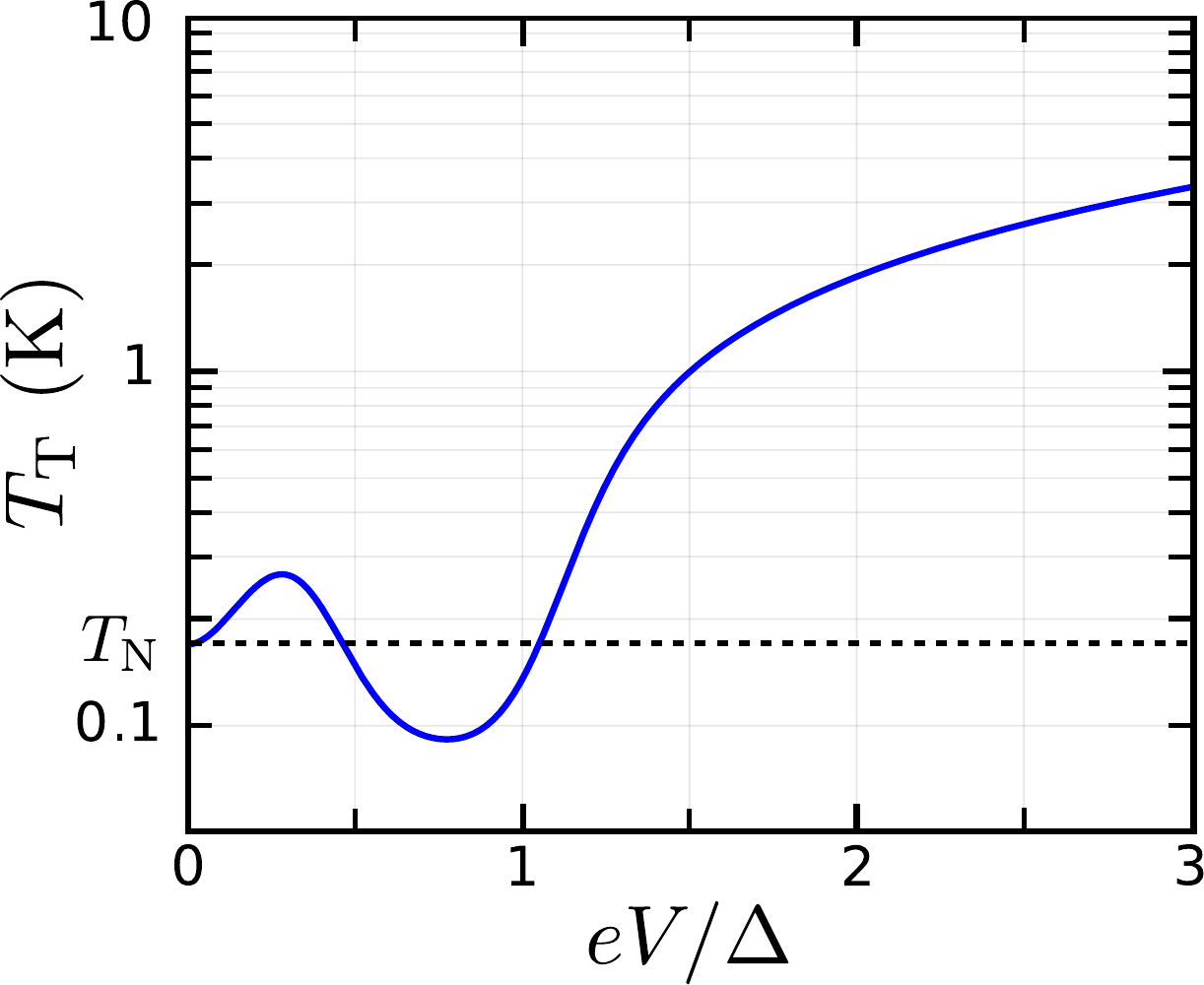}
  \caption{\textbf{Electromagnetic environment}. Calculated temperature $T_{\rm T}$ of the environment as a function of the single-junction bias voltage $V$ for Sample A with the parameters of Table I. At zero bias, the environment temperature equals the electron temperature $T_{\rm N}$ (dashed line) and at high bias it increases linearly. }
\end{figure}
\vspace*{6cm}
\newpage 
\begin{figure}[h]
   \centering
  \includegraphics[height=0.35\linewidth]{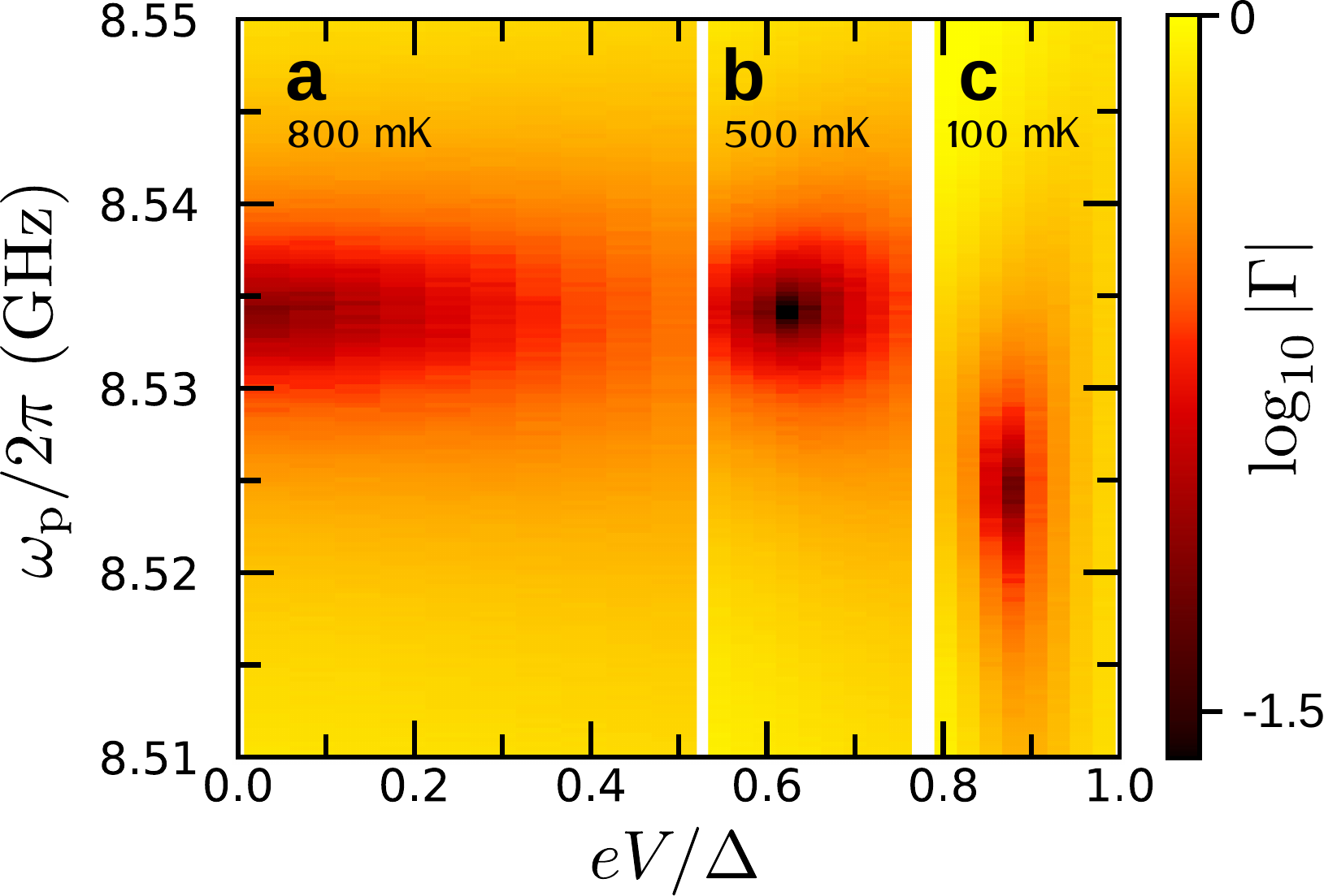}
  \caption{\textbf{Temperature dependence}. \textbf{a--c,} Magnitude of the voltage reflection coefficient $\Gamma=|\Gamma|\exp(-\ii \varphi)$ as a function of the probe frequency $\omega_{\rm p}$ and the single-junction bias voltage $V$ for Sample B at the normal-metal electron temperature $T_{\rm N}=$\SI{800}{\milli\kelvin} in \textbf{a}, \SI{500}{\milli\kelvin} in \textbf{b}, and \SI{100}{\milli\kelvin} in \textbf{c}. }
\end{figure}
\newpage

\begin{figure}[h]
   \centering
  \includegraphics[height=0.35\linewidth]{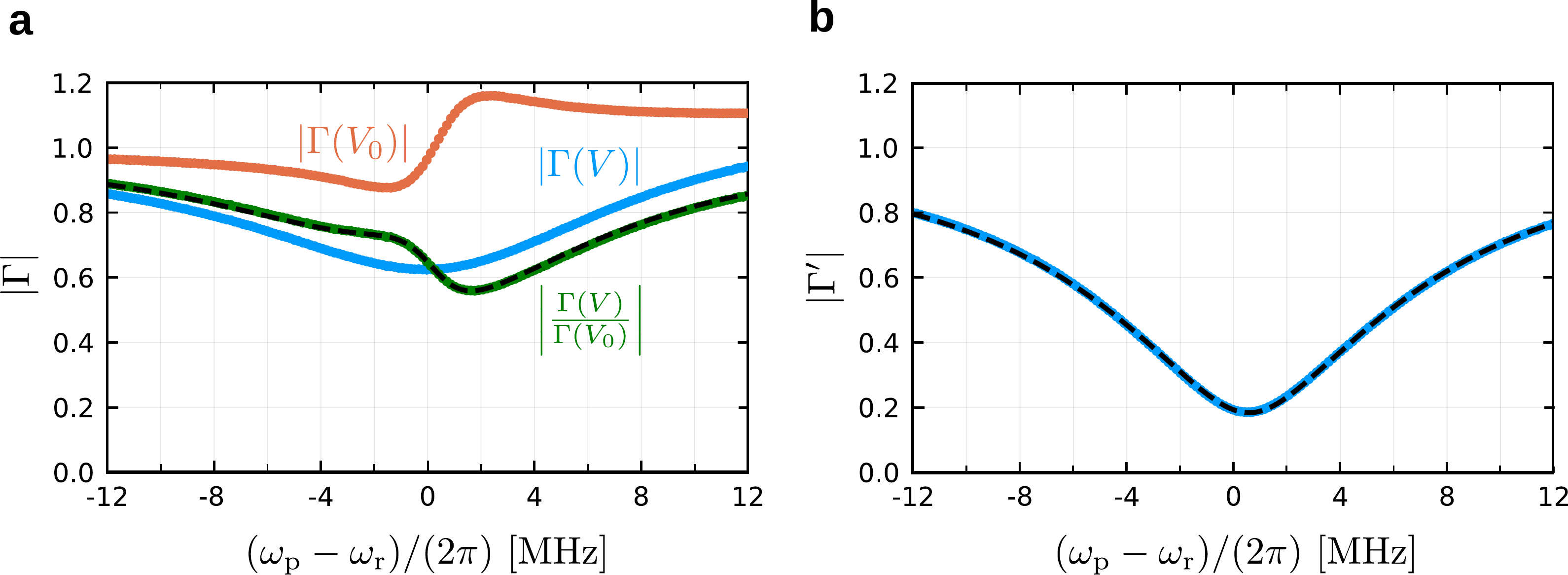}
  \caption{\textbf{Background subtraction from the reflection coefficient data}. \textbf{a,} Magnitude of the measured voltage reflection coefficient $|\Gamma|$ as a function of the probe frequency $\omega_{\rm p}$ for Sample A and two different bias voltages (red and blue colors). In the background subtraction procedure, we divide the measured reflection coefficient at the single-junction bias voltage $V$ (blue) with that at $V_0=0$ (red). From the fit $r=\Gamma(V)/\Gamma(0)$ (dashed line) to the fraction of the two reflection coefficients (green)  ach following equation~(1) of the main text, we extract the parameters of the zero-bias reflection coefficient. We repeat this procedure for every finite bias voltage and finally average the extracted zero-bias parameters, yieldings us $\Gamma'(0)$. \textbf{b,} Magnitude of the background-corrected voltage reflection coefficient $|\Gamma'(V)=|r \Gamma'(0)|$ (blue) and the corresponding fit to equation (1) of the main text (dashed line), where we extract the resonator frequency $\omega_{\rm r}$ and the sum of the external coupling strength~$\gamma_{\rm tr}$ and the excess coupling strength~$\gamma_0$.}
\end{figure}
\newpage

\section*{Supplementary methods 1 | Lamb shift from quasiparticle tunneling}
We present here a derivation of the resonator Lamb shift $\omega_{\rm L}$ caused by the quasiparticle tunneling at the two capacitively coupled superconductor--insulator--normal-metal junctions. The derivation follows the assumptions and guidelines of refs.~\onlinecite{Silveri17prb, IngoldNazarov05}. We apply second-order perturbation theory, since the first-order correction vanishes [see discussion below, after equation~\eqref{eq.mm}]. Thus, we start by introducing the unperturbed Hamiltonians of the electric circuit~$\ot H_0$, the normal-metal electrode~$\ot H_{\rm N}$ and the superconducting leads~$\ot H_{\rm S}$ for the system depicted in Fig.~1a:
\begin{subequations}
\begin{align}
  \ot H_0 &= \sum_{Q_{\rm N}=-\infty}^\infty\sum_{m_{Q_{\rm N}}=0}^\infty \ket{Q_{\rm N},\, m_{Q_{\rm N}}}\bra{Q_{\rm N},\, m_{Q_{\rm N}}} \left(E_{\rm N}Q_{\rm N}^2+\hbar \omega_{\rm r}^0  m_{Q_{\rm N}} \right), \\ \ot H_{\rm N}&=\sum_{\ell \sigma} \varepsilon_\ell \ot d^\dag_{\ell \sigma}\ot d^{}_{\ell \sigma},\\
  \ot H_{\rm S}&=\sum_{k \sigma} (\epsilon_k-eV) \ot c^\dag_{k \sigma}\ot c^{}_{k \sigma} + \sum_k (\Delta_k \ot c^\dag_{k \uparrow}\ot c^\dag_{-k\downarrow} + \Delta^\ast_k \ot c_{k \uparrow}\ot c_{-k\downarrow}). 
\end{align}
\end{subequations}
Hamiltonian $\ot H_0$ corresponds to the charge states of the normal-metal island $\ket{Q_{\rm N}}$, $Q_{\rm N}=0, \pm 1, \ldots$ with the charging energy $E_{\rm N}=e^2/2C_{\rm N}$ and to the resonator with the bare, unnormalized frequency $\omega_{\rm r}^0$ and Fock states $\ket{m}$ where $m=0,1,2,\ldots$ are displaced by the charge of the normal-metal island $\ket{Q_{\rm N},\, m_{Q_{\rm N}}}=\exp\left(-\ii \frac{C_{\rm c}}{C_{\rm N}} Q_{\rm N} \frac{e}{\hbar} \ot \Phi\right)\ket{m}\ket{Q_{\rm N}}$. Here, $\ot \Phi$ refers to the flux of the resonator,  $e$ is the elementary charge, $C_{\rm c}$ is the coupling capacitance, and $C_{\rm N}=C_{\rm c}+C_{\rm{\Sigma m}}$ is the capacitance of the normal-metal island. The energy of the normal-metal (superconducting) lead is represented with $\ot H_{\rm N}$ ($\ot H_{\rm S}$), where the annihilation operator $\ot d_{\ell \sigma}$ ($\ot c_{k \sigma}$) refers to the quasiparticle state with momentum $\ell$ ($k$), energy $\varepsilon_{\ell}$ ($\epsilon_{k}$), and spin $\sigma$. The gap parameter coupling the quasiparticles of the superconductor is denoted with $\Delta_k$. The bias voltage displaces the energy of the superconductor quasiparticles by $eV$.

The perturbation is caused by the quasiparticle tunneling between the normal-metal and superconducting leads represented by the tunneling Hamiltonian~\cite{IngoldNazarov05}
\begin{equation}
  \ot H_{\rm T}=\sum_{\ell k \sigma} \left(T^{}_{\ell k} \ot d_{\ell \sigma}^\dag \ot c^{}_{k \sigma}\ee^{-\ii \frac{e}{\hbar} \ot \Phi_{\rm N}} + T^\ast_{\ell k} \ot d^{}_{\ell \sigma} \ot c^\dag_{k \sigma}\ee^{\ii \frac{e}{\hbar} \ot \Phi_{\rm N}}\right)=\ot \Theta \ee^{-\ii \frac{e}{\hbar} \ot \Phi_{\rm N}} +\ot \Theta^\dag \ee^{\ii \frac{e}{\hbar} \ot \Phi_{\rm N}},
\end{equation}
where $T_{lk}$ is a tunneling matrix element. The perturbation separates to the electronic and electric parts. The electronic part $\ot \Theta$ describes the quasiparticle transitions and the electric part $\ee^{\pm \ii \frac{e}{\hbar} \ot \Phi_{\rm N}}$ describes the associated transitions in the state of the electromagnetic degrees of freedom, namely, the displacement of the island charge $\ot Q_{\rm N}\to \ot Q_{\rm N}\pm e$, where $\ot \Phi_{\rm N}$ and $\ot Q_{\rm N}$ refer to the flux and charge of the normal-metal island. This tunneling Hamiltonian corresponds the transitions through one of the superconductor--insulator--normal-metal junctions. Small junction conductance $G_{\rm T} \lesssim$~\SI{100}{\micro\siemens} implies that the probability for co-tunneling is negligible small and tunneling at the two junctions can be considered separately. We add the contribution of the other junction below. 

The energy level shift $\hbar \delta_{\eta}$ by the second order time-independent perturbation theory is 
\begin{equation}
  \hbar\delta_\eta=E_{\eta}-E_\eta^0=-\sum_{\eta'\neq \eta}\frac{|\braket{\eta^\prime|\ot H_{\rm T}|\eta}|^2}{E_{\eta^\prime}-E_\eta}, \label{eq.eshift}
\end{equation}
where $\ket{\eta}=\ket{Q_{\rm N}, m_{Q_{\rm N}}, \ell, k}$ is a notation for the combined state the unperturbed system with the total energy $E_\eta=E_{\nmd} Q_{\rm N}^2+\hbar \omega_{\rm r}^0 m+\varepsilon_l+\epsilon_k$. Since $\ee^{\pm \ii \frac{e}{\hbar} \ot \Phi_{\rm N}}$ is the displacement operator of the island charge, it yields the matrix element of the electric circuit
\begin{align}
  \braket{m_{Q'_{\rm N}},\,Q'_{\rm N}|\ee^{\pm \ii \frac{e}{\hbar} \ot \Phi_{\rm N}}|Q^{}_{\rm N},\, m_{Q^{}_{\rm N}}}=\delta_{Q'_{\rm N}, Q^{}_{\rm N}\pm 1}\braket{m'_{Q_{\rm N}\pm 1}|m^{}_{Q_{\rm N}}} =\delta_{Q'_{\rm N},Q^{}_{\rm N}\pm 1}\braket{m'_{\pm 1}|m^{}_{}}=\delta_{Q'_{\rm N},Q_{\rm N}\pm 1} M_{mm'}, \label{eq.mees}
\end{align}
where $M_{mm'}$ is the matrix element between the charge-shifted Fock states of the resonator~\cite{Silveri17prb, Catelani11}
\begin{align}
|M_{m m^\prime}|^2 =\left|\int\psi_{m^\prime}^\ast \left(q\pm\alpha e\right) \psi_m\left(q\right)  \Dn{q}\right|^2 = \ee^{-\rho} \rho^\ell \frac{m^\prime !}{m!} \left[L^{\ell}_{m^\prime }(\rho)\right]^2. \label{eq.mm}
\end{align}
Here $\psi_m(q)=\braket{q|m}$ are the resonator energy eigenstates represented in the charge $q$ basis, $\rho=\pi \frac{C^2_{\rm c}}{C^2_{\rm N}}\frac{Z_{\rm r}}{R_{\rm K}}$ is an interaction parameter expressed in terms of the characteristic impedance $Z_{\rm r}$ of the resonator and the von Klitzing constant~$R_{\rm K}=h/e^2$, and $L^{\ell}_{m^\prime }(\rho)$ denote the generalized Laguerre polynomials~\cite{Stegun}. Note that since the matrix element of equation~\eqref{eq.mees} involves a charge shift, the energy shift by the first order perturbation theory vanishes.

Given the matrix element of the electric circuit of equation~\eqref{eq.mees}, the frequency shift of equation~\eqref{eq.eshift} reduces to
\begin{align}
  \hbar \delta_{\eta}= -\sum_{m', \ell', k'} |M_{mm'}|^2\bigg[ &\frac{|\braket{\ell' k'|\ot \Theta^\dag|\ell k}|^2}{E_{\rm N}(1-2Q_{\rm N})+\hbar \omega_{\rm r}^0 (m'-m)+E_{\ell^\prime k^\prime}-E_{\ell^{} k^{}}}\notag \\ & \quad\qquad+\frac{|\braket{\ell' k'|\ot \Theta|\ell k}|^2}{E_{\rm N}(1+2Q_{\rm N})+\hbar \omega_{\rm r}^0 (m'-m)+E_{\ell^\prime k^\prime}-E_{\ell^{} k^{}}} \bigg]. \label{eq.eshift.q}
\end{align}
Since the charging energy $E_{\rm N}=e^2/2(C_{\rm c}+C_{\rm \Sigma m}) \sim h \times$ \SI{10}{\mega\hertz} is the smallest of the relevant energy scales of the setup ($\Delta$, $\hbar \omega_{\rm r}^0$, and $k_{\rm B}T_{\rm N}$), we can expand equation~\eqref{eq.eshift.q} in $2E_NQ_{\rm N}$ and average over the symmetric charge state distribution $p_{Q_{\rm N}}$~\cite{Silveri17prb}. Furthermore, we trace over the state of the normal-metal and the superconducting leads. The result is  
\begin{align}
  \hbar \widetilde{\delta}_{m} &= \sum_{\ell,k} \sum_{Q_{\rm N}} p_{Q_{\rm N}} \delta_{Q_{\rm N} m \ell k}\notag \\
                               &= - \sum_{m'}\sum_{\ell,\ell'}\sum_{k,k'} |M_{mm'}|^2 \left[\frac{|\braket{\ell' k'|\ot \Theta^\dag|\ell k}|^2}{E_{\rm N}+\hbar \omega_{\rm r}^0 (m'-m)+E_{\ell^\prime k^\prime}-E_{\ell k}} +\frac{|\braket{\ell' k'|\ot \Theta|\ell k}|^2}{E_{\rm N}+\hbar \omega_{\rm r}^0 (m'-m)+E_{\ell^\prime k^\prime}-E_{\ell k}} \right].
\end{align}
From equation~\eqref{eq.mm} we expand $|M_{mm'}|^2$ up to the first order in $\rho$, which is justified by the typical experimental values $\rho \approx 0.001$, resulting in $|M_{m,m}|^2=1- (1+2m)\rho+\mathcal O(\rho^2)$,  $|M_{m,m+1}|^2 = \rho (m+1)+\mathcal O(\rho^2)$, $|M_{m,m-1}|^2=\rho m \mathcal + O(\rho^2)$, and $|M_{m,m\pm s}|^2=\mathcal O (\rho^s)$ for $s \ge 2$. By taking into account the terms up to the first order in $\rho$, we obtain that the Lamb shift of the harmonic oscillator $ \omega_{\rm{L}}   =  \widetilde{\delta}_{m+1}-\widetilde{\delta}_{m}$ is 
\begin{align}
  \omega_{\rm{L}}   = & \frac{\rho}{\hbar} \sum_{\ell,\ell'} \sum_{k,k'} \sum_{\tau = \pm 1} \left( \frac{|\braket{\ell' k'|\ot \Theta^\dag|\ell k}|^2+|\braket{\ell' k'|\ot \Theta|\ell k}|^2}{E_{\rm N}+E_{\ell^\prime k^\prime}-E_{\ell k}} - \frac{|\braket{\ell' k'|\ot \Theta^\dag|\ell k}|^2+|\braket{\ell' k'|\ot \Theta|\ell k}|^2}{E_{\rm N}+\tau \hbar \omega_{\rm r}^0+E_{\ell^\prime k^\prime}-E_{\ell k}}
                        \right ) \notag \\            
                                 =& \omega^{\rm{el}}_{\rm {L}}+\omega^{\rm{ph}}_{\rm{L}}, \label{eq.lamb01}
\end{align}
where we have denoted the first and the second part by $\omega^{\rm{el}}_{\rm{L}}$ and $\omega^{\rm{ph}}_{\rm{L}}$, respectively. The part $\omega^{\rm{ph}}_{\rm{L}}$ originates from the photon-assisted transitions, whereas the part $ \omega^{\rm{el}}_{\rm{L}}$ originates from the elastic transitions and can be expressed as $\omega^{\rm{el}}_{\rm{L}}=- \lim_{\omega_{\rm r}^0 \to 0} \omega^{\rm{ph}}_{\rm {L}}(\omega_{\rm r}^0)$. Thus we begin by simplifying $\omega^{\rm{ph}}_{\rm {L}}$. 

To calculate the matrix elements of the quasiparticle transitions $|\braket{\ell' k'|\ot \Theta|\ell k}|$ in equation~\eqref{eq.lamb01}, we proceed similarly as in refs.~\onlinecite{IngoldNazarov05, Silveri17prb} by expressing the matrix elements in terms of the Fermi functions of the normal-metal $f_{\rm N}(\varepsilon)$ and superconducting leads $f_{\rm S}(\epsilon)$ and the normalized quasiparticle density of the states in the superconductor $n_{\rm S}(\epsilon)$. Furthermore, we assume that the tunneling matrix elements $T_{\ell k}$ are approximately constant around the Fermi energies and their effect can be expressed with the junction conductance $G_{\rm T}$. The result is
\begin{align}
  \omega^{\rm{ph}}_{\rm{L}}=-\frac{\rho }{ 2\pi} R_{\rm K} G_{\rm T} \sum_{\tau = \pm 1} \bigg\{ &\frac{1}{h} \textrm{PV}\iint_{-\infty}^{\infty} \Dn{\epsilon_k} \Dn{\varepsilon_\ell} \frac{n_{\rm S}(\epsilon_k)\left[1-f_{\rm S}(\epsilon_k)\right]f_{\rm N}(\varepsilon_{\ell})}{E_{\nmd}+\tau \hbar \omega_{\rm r}^0-eV + \epsilon_k-\varepsilon_{\ell}} \notag \\
                         &\quad+\frac{1}{h} \textrm{PV}\iint_{-\infty}^{\infty} \Dn{\epsilon_k} \Dn{\varepsilon_\ell} \frac{n_{\rm S}(\epsilon_k)f_{\rm S}(\epsilon_{k})\left[1-f_{\rm N}(\varepsilon_\ell)\right]}{E_{\nmd}+\tau \hbar \omega_{\rm r}^0+e V + \varepsilon_{\ell}-\epsilon_k} \bigg \}, \label{eq.lambpv}
\end{align}
where $\textrm{PV}$ denotes the Cauchy principal value integration. Next we consider also the other normal-metal--insulator--superconductor junction in the construction. By assuming that the temperatures of the all the leads in the construction are identical ($f_{\rm N}=f_{\rm S}$) and that the junction conductances are identical, it follows that the only difference in the derivation of the Lamb shift by the other junction is that the bias voltage is opposite $eV \to -eV$ in equation~\eqref{eq.lambpv}. We sum up the contributions from both junctions resulting in
\begin{align}
  \omega^{\rm{ph}}_{\rm{L}} &=-\frac{\rho }{2\pi}R_{\rm K}G_{\rm T}\sum_{\sigma=\pm 1} \sum_{\tau = \pm 1}\bigg \{ \frac{1}{h} \textrm{PV}\iint_{-\infty}^{\infty} \Dn{\epsilon_k} \Dn{\varepsilon_\ell} \frac{n_{\rm S}(\epsilon_k)\left[1-f(\epsilon_k)\right]f(\varepsilon_{\ell}-\sigma eV-\tau \hbar \omega_{\rm r}^0+E_{\nmd})}{\epsilon_k-\varepsilon_{\ell}}\notag \\
                       & \qquad \qquad \quad \qquad \qquad \qquad \ -\frac{1}{h}\textrm{PV}\iint_{-\infty}^{\infty} \Dn{\epsilon_k} \Dn{\varepsilon_\ell} \frac{n_{\rm S}(\epsilon_k)f(\epsilon_{k})\left[1-f(\varepsilon_\ell-\sigma eV-\tau \hbar \omega_{\rm r}^0-E_{\nmd})\right]}{\epsilon_k-\varepsilon_{\ell}} \bigg\} \label{eq.lambp}.
\end{align}
Next we utilize the notion of the normalized rate of forward  $\normrateright(E)$  and backward $\normrateleft(E)$  quasiparticle tunneling for a junction with tunneling resistance $R_{\rm T}$ equal to the von Klitzing constant $R_{\rm K}=h/e^2$
\begin{align}
  \normrateright(E) &= \frac{1}{h}\int_{-\infty}^\infty \Dn{\varepsilon}\, n_{\rm S}(\varepsilon) [1-f(\varepsilon)]f(\varepsilon-E), & 
  \overleftarrow{F}(E) & =\frac{1}{h}\int_{-\infty}^\infty \Dn{\epsilon}\, n_{\rm S}(\varepsilon) f(\varepsilon)  \left[1-f\left(\varepsilon-E \right)\right]. \label{eq:PEback}
\end{align}
These rates obey the symmetry $\normrateright(-E)=\normrateleft(E)$ resulting in that equation~\eqref{eq.lambp} can be expressed in a simple form
\begin{align}
  \omega^{\rm{ph}}_{\rm{L}}&=-\frac{\rho}{\pi } R_{\rm K}G_{\rm T}\sum_{\sigma=\pm 1} \sum_{\tau=\pm 1}   \textrm{PV}\int_{-\infty}^\infty \Dn{\epsilon} \frac{ \normrateright(\epsilon+\sigma eV+\tau \hbar \omega_{\rm r}^0-E_{\rm N})}{\epsilon} \notag \\
  &=-\pi \frac{C^2_{\rm c}}{C^2_{\rm N}} Z_{\rm r} G_{\Sigma} \sum_{\sigma=\pm 1} \sum_{\tau=\pm 1} \left[\textrm{PV}\int_{0}^\infty\frac{\Dn{\omega}}{2\pi} \frac{  \tau \normrateright(\sigma eV+\tau \hbar \omega-E_{\rm N})}{\omega-\omega_{\rm r}^0}+\textrm{PV}\int_{0}^\infty \frac{\Dn{\omega}}{2\pi} \frac{ \tau \normrateright(\sigma eV+\tau \hbar \omega-E_{\nmd})}{\omega+\omega_{\rm r}^0}\right], \label{eq.lamb.almost}                      
\end{align}
where the total conductance is $G_{\Sigma}=2G_{\rm T}$. Furthermore, by expressing equation~\eqref{eq.lamb.almost} in terms of the coupling strength $\gamma_{\rm T}$ of the effective electromagnetic environment derived in ref.~\onlinecite{Silveri17prb},
\begin{equation}
  \gamma_{\rm T}(V, \omega)=\pi \frac{C^2_c}{C_{\rm N}^2} Z_{\rm r} G_{\Sigma} \sum_{\sigma=\pm 1}\sum_{\tau=\pm 1} \tau \normrateright(\sigma eV+\tau \hbar \omega-E_{\nmd}),
\end{equation}
we finally arrive with the total Lamb shift by both the photon-assisted and elastic tunneling transitions $\omega_{\rm{L}}=\omega^{\rm{ph}}_{\rm{L}}+\omega^{\rm{el}}_{\rm{L}}= \omega^{\rm{ph}}_{\rm{L}}(\omega_{\rm r}^0) - \lim_{\omega_{\rm r}^0 \to 0} \omega^{\rm{ph}}_{\rm {L}}(\omega_{\rm r}^0)$
\begin{equation}
\omega_{\rm{L}}\left (V, \omega_{\rm r}^0 \right)  =  - \textrm{PV}\int_{0}^\infty \frac{\Dn{\omega}}{2\pi}\left[ \frac{\gamma_{\rm T}(V, \omega)}{\omega-\omega_{\rm r}^0}+\frac{\gamma_{\rm T}(V, \omega)}{\omega+\omega_{\rm r}^0}-2\frac{\gamma_{\rm T}(V, \omega)}{\omega}\right]. \label{eq:lamb}
\end{equation}

\section*{Supplementary Methods 2 --- Lamb shift from the Kramers-Kronig relations}
In general, causality imposes restrictions on  linear-response coefficients in frequency space, referred to as Kramers--Kronig relations. In particular, the real and the imaginary part of any physical admittance $Y(\omega)$ are related by the Hilbert transform which reads
\begin{equation}\label{eq:kk}
  \operatorname{Im} Y(\omega) = 
  \textrm{PV}\int_{-\infty}^{\infty}\!\frac{\textrm{d}\omega}{\pi}'\, \frac{\operatorname{Re}
    Y(\omega')}{\omega' - \omega} =
    \textrm{PV}\int_0^\infty\!\frac{\textrm{d}\omega'}{\pi}\, \operatorname{Re}
    Y(\omega') \left(\frac1{\omega' - \omega} - \frac1{\omega' + \omega}  \right).
\end{equation}
As above, we model the resonator without coupling to the environment as a harmonic oscillator realized by a parallel $LC$circuit with voltage $V$, characteristic impedance $Z_r$ and resonance frequency $\omega^0_{\rm r}$. The environment adds a small shunting admittance with $|Y| Z_{\rm r} \ll 1$. The Kirchhoff current rule reads
\begin{equation}
  V(\omega) \left[ \frac{\ii[\omega^2 -(\omega_{\rm r}^0)^2]}{Z_{\rm r} \omega^0_{\rm r} \omega} +
    Y(\omega)\right ] = 0
\end{equation}
where the first term in the bracket is the admittance of the resonator. The modes of the system thus correspond to the roots of the term in the bracket. Given the fact that the shunt is small, we can obtain an approximate expression for the position  (frequency and damping) of the eigenmode including the environment. In particular, we parameterize the root by $\omega =  \omega^0_{\rm r} + \omega_{\rm L} + \ii \gamma_{\rm T}/2$ with $\omega_{\rm L}$ the Lamb shift and $\gamma_{\rm T}$ the coupling strength of the environment. As a result, we obtain a relation of the admittance to the relevant quantities of the form
\begin{align}\label{eq:kk2}
  \omega_{\rm L}(\omega^0_{\rm r}) &= -\frac{\omega^0_{\rm r} Z_{\rm r}}2 \operatorname{Im}Y(\omega^0_{\rm r}), & \gamma_{\rm T}(\omega^0_{\rm r})&= \omega^0_{\rm r} Z_{\rm r} \operatorname{Re} Y(\omega^0_{\rm r}).
\end{align}
With these relations, we obtain from equation \eqref{eq:kk} the Kramers--Kronig-type relation between the coupling strength and the Lamb shift
\begin{equation}\label{eq:lamb2}
  \omega_{\rm L}(\omega^0_{\rm r}) = - \textrm{PV}\int_0^\infty\!\frac{\Dn{\omega}}{2\pi}\,  \frac{\gamma_{\rm T}(\omega)}{\omega} \left( \frac{\omega^0_{\rm r}}{\omega - \omega^0_{\rm r} }- \frac{\omega^0_{\rm r}}{\omega + \omega^0_{\rm r}}\right)\,.
\end{equation}
Note that this expression coincides with equation~\eqref{eq:lamb}. 

However, we stress that a frequency-independent coupling strength $\gamma_\textrm{T}$ rise to a vanishing Lamb shift in equation~\eqref{eq:lamb2} and similarly a frequency-independent Lamb shift yields a vanishing contribution to the coupling strength. Thus equations~\eqref{eq:kk} and~\eqref{eq:kk2} are only valid up to frequency-independent shifts. These frequency-independent shifts are identical to the static Lamb shifts we consider in the main article in addition to the dynamic Lamb shift given by equations~\eqref{eq:lamb} and~\eqref{eq:lamb2}. If a static shift is independent of the bias voltage, it is not resolved in the experiment, and hence we only consider static shifts linear in the coupling strength, \textit{i.e.}, the lowest-order static corrections to the dynamic shift.